\documentclass[twocolumn,prl,twocolumn,superscriptaddress]{revtex4}
\usepackage[latin9]{inputenc}
\usepackage{color}
\usepackage{amsmath}
\usepackage{amssymb}
\usepackage{graphicx}
\usepackage{tikz}

\usepackage{mathrsfs} 
\usepackage{float}
\usepackage{bbm}

\usepackage{epsfig} 
\usepackage{verbatim}
\usepackage{array}
\usepackage{setspace}
\usepackage{ulem}

\usepackage[unicode=true,
 bookmarks=true,bookmarksnumbered=false,bookmarksopen=false,
 breaklinks=false,pdfborder={0 0 1},backref=false,colorlinks=true]
 {hyperref}
\hypersetup{
 linkcolor=magenta, urlcolor=blue, citecolor=blue, pdfstartview={FitH}, hyperfootnotes=false, unicode=true}

\makeatletter
\@ifundefined{textcolor}{}
{%
 \definecolor{BLACK}{gray}{0}
 \definecolor{WHITE}{gray}{1}
 \definecolor{RED}{rgb}{1,0,0}
 \definecolor{GREEN}{rgb}{0,1,0}
 \definecolor{BLUE}{rgb}{0,0,1}
 \definecolor{CYAN}{cmyk}{1,0,0,0}
 \definecolor{MAGENTA}{cmyk}{0,1,0,0}
 \definecolor{YELLOW}{cmyk}{0,0,1,0}
}

\usepackage{amsfonts}\usepackage{tabularx}\usepackage{dcolumn}\usepackage{bm}\usepackage{graphicx}\usepackage{epstopdf}

\hypersetup{urlcolor=blue}

\makeatother

\newcolumntype{C}[1]{>{\centering\arraybackslash$}p{#1}<{$}}
\newcommand{\rr}{\bm{r}}

\begin{document}

\title{Magic angle for barrier-controlled double quantum dots}

\author{Xu-Chen Yang}
\affiliation{Department of Physics, City University of Hong Kong, Tat Chee Avenue, Kowloon, Hong Kong SAR, China}
\affiliation{City University of Hong Kong Shenzhen Research Institute, Shenzhen, Guangdong 518057, China}
\author{Xin Wang}
\email{x.wang@cityu.edu.hk}
\affiliation{Department of Physics, City University of Hong Kong, Tat Chee Avenue, Kowloon, Hong Kong SAR, China}
\affiliation{City University of Hong Kong Shenzhen Research Institute, Shenzhen, Guangdong 518057, China}
\date{\today}

\begin{abstract}

We show that the exchange interaction of a singlet-triplet spin qubit confined in double quantum dots, when being controlled by the barrier method, is insensitive to a charged impurity lying along certain directions away from the center of the double-dot system. These directions differ from the polar axis of the double dots by the magic angle, equaling $\arccos\left(1/\sqrt{3}\right)\approx 54.7^\circ$, a value previously found in atomic physics and nuclear magnetic resonance.  This phenomenon can be understood from an expansion of the additional Coulomb interaction created by the impurity, but also relies on the fact that the exchange interaction solely depends on the tunnel coupling in the barrier-control scheme. Our results suggest that for a scaled-up qubit array, when all pairs of double dots rotate their respective polar axes from the same reference line by the magic angle, cross-talks between qubits can be eliminated, allowing clean single-qubit operations.
While our model is a rather simplified version of actual experiments, our results suggest that it is possible to minimize unwanted couplings by judiciously designing the layout of the qubits.

\end{abstract}

\maketitle

\section{Introduction}
Despite their simplicity, the semiconductor double quantum dots are among the most extensively studied in physics \cite{vanderWiel.02}. For example, they are platforms to study interesting transport phenomena including the Coulomb blockade \cite{Waugh.95,Matveev.96} and the Kondo effect \cite{Pohjola.01}. Recently, they play an important role in the search of a viable physical system to host a quantum computer \cite{nielsen2010quantum}, thanks to the technological advance in fabrication, manipulation and measurement of these devices \cite{Petta.05, Bluhm.10b, Barthel.10,Maune.12, Pla.13,Muhonen.14,Kim.14,Kawakami.16} and their potential of scalability \cite{Taylor.05}. While there are many ways to encode a qubit using either charge \cite{Gorman.05} or spin states \cite{Loss.98, DiVincenzo.00, Levy.02,Shi.12} of electrons confined in the quantum dots, the singlet-triplet spin qubit is among the most successful ones because it is the simplest type that can be controlled solely electrostatically \cite{Petta.05, Foletti.09,Bluhm.10c,vanWeperen.11,Maune.12, Shulman.12, Wu.14, Nichol.16}. The key control parameter is the Heisenberg exchange interaction between the two spins, which can be varied either by changing the relative energy of the two dots (``tilt control'') \cite{Petta.05}, or by raising and lowering the central potential barrier with the two dots kept leveled in energy (``barrier control'') \cite{Reed.16, Martins.16}. Since the barrier control essentially operates the qubit near a ``sweet spot'' where the charge noise \cite{Hu.06, Wardrop.14, Thorgrimsson.16} is substantially suppressed \cite{Yang.17b}, this method holds great promise for high-fidelity universal qubit manipulation  \cite{Zhang.17}.

The ``magic angle'', defined as $\theta_m=\arccos\left(1/\sqrt{3}\right)$ ($\approx 54.7^\circ$), appears in many fields of physics and related sciences. 
It has been discovered early on in atomic physics by measuring the polarization of the resultant radiation when the mercury vapor is illuminated by a polarized light \cite{Wood.23}. It has been found that when a magnetic field was applied at $\theta_m$ with respect to the polarization axis of the activating light, the resultant radiation appears unpolarized \cite{Eldridge.24,Van.25}. This interesting effect has since been rediscovered in various other contexts of atomic physics \cite{Gunn.71,Deech.75,Sansonetti.11}.
 In solid-state nuclear magnetic resonance spectroscopy, it was discovered in an effort to improve the signal-to-noise ratio, that if a solid sample is spun at this angle relative to the applied magnetic field, dipolar interactions between nuclei are suppressed and the observed spectral lines are much sharper \cite{Lowe.59,Andrew.59,Dreitlein.61}. This has created the technique of magic-angle spinning \cite{Kessemeier.67,Hennel.05}, which has been used subsequently in chemistry \cite{Handel.03}, medicine \cite{Bydder.07}, and quantum computation \cite{Ding.01}. The magic angle is also of theoretical interest: for example, it has been shown to significantly influence the quantum state transfer along a Heisenberg spin chain \cite{Lu.09}. While the magic angle arises in various physical situations, it can be understood mathematically as the root of the second-order  Legendre polynomial, which arises for example from the multipole expansion of $1/r$. Any interaction dependent on it shall therefore vanish at this angle.

In this paper we shall show that the exchange interaction of a singlet-triplet qubit, while being barrier-controlled, is insensitive to an impurity situated along the direction which is precisely at the magic angle apart from the polar axis of the double dots \cite{polar}. This interesting phenomenon stems essentially from the multipole expansion of the additional Coulomb repulsion created by the impurity, but also relies on the fact that the exchange interaction solely depends on the tunnel coupling in the barrier-control scheme. While this finding can be easily generalized to multiple impurities, the most interesting implication is for a series of double dots: when each pair of the double dots has its polar axis rotated from the same reference line by the magic angle, the operation using exchange interaction shall not be affected by other dots in the array.

\section{Model}

\begin{figure}[t]
(a)\centering\includegraphics[width=0.6\columnwidth]{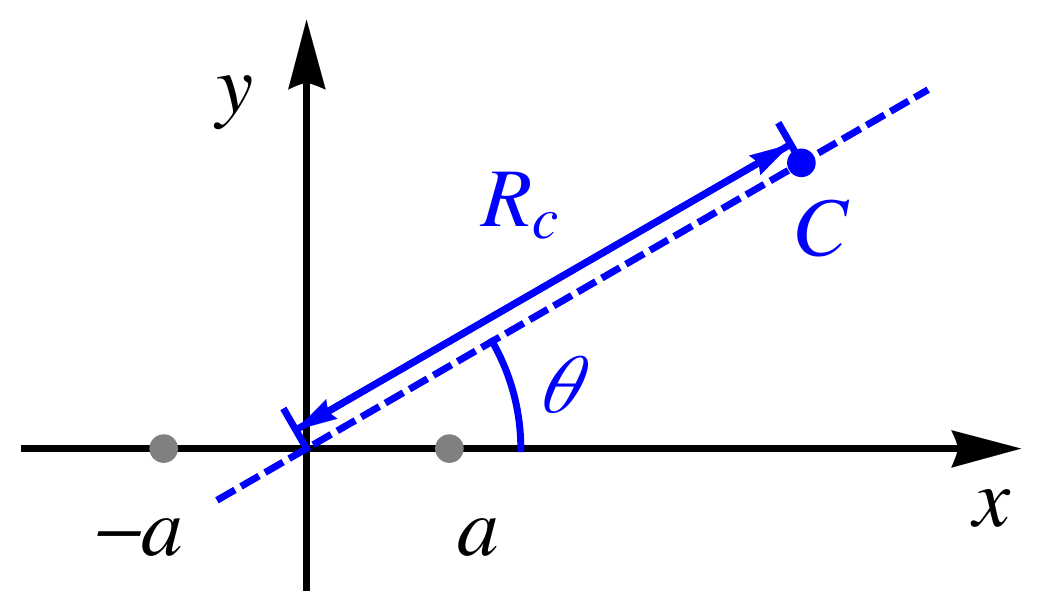}
\hspace{\fill}
(b)\centering\includegraphics[width=0.6\columnwidth]{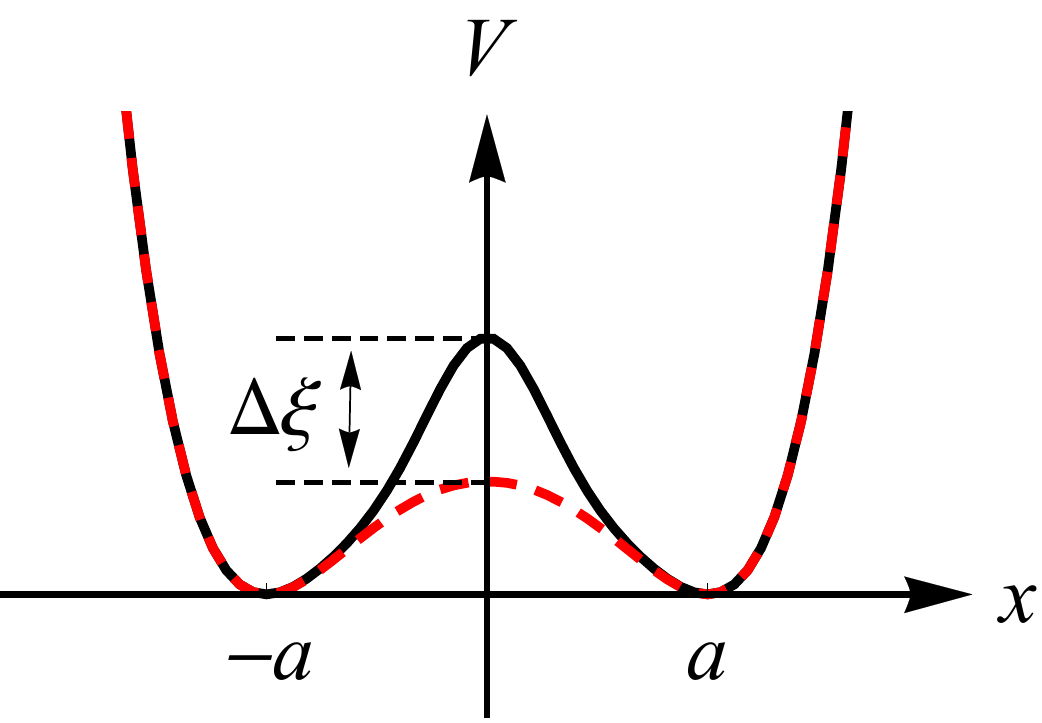}
\caption{(a) Schematic diagram showing the double quantum dots locating at $(\pm a,0)$ together with a charged impurity $C$ at $\bm{R}_c=(R_c\cos\theta,R_c\sin\theta)$.
(b) Schematic double-well confinement potential of a double-quantum-dot system under barrier control. The barrier control method changes the height of the central potential barrier (here by $\Delta\xi$), which subsequently varies the exchange energy. }
\label{fig:dqd}
\end{figure}

We consider a double-quantum-dot system in the $xy$ plane lying along the $x$ direction [Fig.~\ref{fig:dqd}(a)]. When two electrons are allowed in the system, its Hamiltonian can be written in a second-quantized form as \cite{Burkard.99,Yang.11,Wang.11}:
\begin{equation}
\begin{split}
H=&-\mu_1(n_{1\uparrow}+n_{1\downarrow})-\mu_2(n_{2\uparrow}+n_{2\downarrow})+U_1n_{1\uparrow}n_{1\downarrow}\\
&+U_2n_{2\uparrow}n_{2\downarrow}+U_{12}(n_{1\uparrow}+n_{1\downarrow})(n_{2\uparrow}+n_{2\downarrow})\\
&+t\sum_\sigma\left(c^\dagger_{1\sigma}c_{2\sigma}+\mathrm{H.c.}\right),\label{eq:firstHam}
\end{split}
\end{equation}
where $c_{i\sigma}^\dagger$ creates an electron with spin $\sigma$ on the $i{\rm th}$ dot $(i=1,2)$, $-\mu_i$ is the energy of the electron in the the $i{\rm th}$ dot, $U_{i}$ and $U_{12}$ are on-site and inter-site Coulomb interactions, and $t$ is the tunnel coupling between the two dots.

A singlet-triplet qubit is formed when each dot is occupied by one electron. The key control parameter is the Heisenberg exchange interaction between the two electrons, which can be expressed effectively as \cite{Li.12,Yang.17a, Yang.17b}
\begin{equation}
J\approx\frac{2t^2}{\Delta U+\varepsilon}+\frac{2t^2}{\Delta U-\varepsilon},
\label{eq:jeff}
\end{equation}
where $\varepsilon=\mu_2-\mu_1$ (the ``detuning''), $\Delta U=U_1-U_{12}$ and $U_1=U_2$ in this work. Once the form of the confinement potential is known, all parameters above can be readily calculated using  the molecular orbital theory, for example the configuration interaction method in conjunction with the Hund-Mulliken approximation \cite{Burkard.99}. Fig.~\ref{fig:dqd}(b) shows an example of the confinement potential for two dots centering at $(\pm a,0)$ in the form prescribed in \cite{Yang.17b}. While its detailed form is complicated and we would simply refer the reader to Eq.~(5) in \cite{Yang.17b} or Appendix~\ref{appxA}, we note that both wells are well approximated by a harmonic oscillator potential with energy level spacing $\hbar\omega_0$, and the height of the central barrier is characterized by $\xi$. $\hbar\omega_0$, $\xi$ and $a$ exclusively define the potential and the barrier control is done by changing $\xi$ (while $\mu_1=\mu_2$). The tunnel coupling $t$ depends crucially on the height of the central barrier  $\xi$: $t$ decreases almost exponentially as the barrier is raised, but will rapidly increase when the barrier is lowered \cite{Simmons.09}. Both $t$ and $\varepsilon$  significantly affect the amplitude of the exchange interaction $J$. The traditional tilt control method amounts to changing the detuning $\varepsilon$, while the barrier control is equivalent to changing $t$ while keeping $\varepsilon=0$. $\varepsilon=0$ is a ``sweet spot''  for charge noise, making the barrier-control method superior.

A charged impurity adds additional Coulomb interaction to the electrons inside the quantum dots, changes their energy levels and consequently the exchange interaction, causing the ``charge noise''. Fig.~\ref{fig:dqd}(a) shows an impurity $C$ situated at a distance $R_c$ away from the center of the double dots,  lying in a direction $\theta$ with respect to the $x$-axis. In \cite{Yang.17b} we have studied in detail the relative charge noise, defined as the shift in the exchange interaction divided by its magnitude, which can be written using Eq.~\eqref{eq:jeff} as
\begin{equation}
\begin{aligned}
\frac{\delta J}{J}&=\frac{1}{J}\frac{\partial J}{\partial t}\delta t+\frac{1}{J}\frac{\partial J}{\partial\varepsilon}\delta\varepsilon\\
&=\frac{2}{t}\delta t+\frac{2\varepsilon}{\Delta U^2-\varepsilon^2}\delta\varepsilon.
\label{eq:dJfunct}
\end{aligned}
\end{equation} 
Both $\delta t$ and $\delta\varepsilon$ depend on the position of the impurity. Nevertheless, for a barrier-controlled singlet-triplet qubit, $\varepsilon=0$ and the relative charge noise only depends on $\delta t$. We shall show, under the Hund-Mulliken approximation, that $\delta t$ shall vanish when the impurity is along a direction that differs from the polar axis of the double dots (in this case, $\hat{x}$) by the magic angle, i.e. $\theta=\theta_m$. As far as the exchange interaction is concerned, the double dots seem ``blind'' to such an impurity. This has very interesting implications to be detailed later in this paper.

\section{Results}

We follow the formalism already established in \cite{Yang.17b}. An impurity changes the Hubbard parameters as $\mu_1\rightarrow\mu_1-Z_{t1}$, $\mu_2\rightarrow\mu_2-Z_{t2}$, and $t\rightarrow t-Z_{t12}$ (cf. Eq.~(10) in \cite{Yang.17b}). However, shifts in $\mu_1$ and $\mu_2$ are unimportant for our purpose and the key term is $Z_{t12}$. Denoting the wave function for the electron in the $i$th dot as $\psi_i$, and the additional Coulomb interaction at $\bm{r}$ caused by an impurity (with charge $-e$) at $\bm{R}_c$ as $Z=e^2/(4\pi\kappa|\bm{r}-\bm{R}_c|)$, we have $Z_{t12}=\langle\psi_1|Z|\psi_2\rangle$. (For the convenience of discussion we name $\bm{R}_c$ as the ``impurity vector''.) Under the Hund-Mulliken approximation, we express the electron wave functions as linear superpositions of Fock-Darwin states $\phi_i$:
\begin{equation}
\left(
\begin{array}{cccc}
\psi_1\\
\psi_2
\end{array}
\right)=
\left(
\begin{array}{cccc}
m&n\\
n&m
\end{array}
\right)\left(
\begin{array}{cccc}
\phi_1\\
\phi_2
\end{array}
\right),
\end{equation}
and the orthogonality requires that $mn={g}/[2(g^2-1)]$ and $m^2+n^2={1}/(1-g^2)$, where $m$, $n$ are real numbers, $g=\langle\phi_1|\phi_2\rangle=\exp(-a^2/a_B^2)$ and  $a_B$ is the corresponding Fock-Darwin radius \cite{Yang.17b}. $Z_{t12}$ can then be written as
\begin{equation}
Z_{t12}=mnz_{t11}+m^2z_{t12}+n^2z_{t21}+mnz_{t22},\label{eq:bigZt12}
\end{equation}
where $z_{tij}=\langle\phi_i|Z|\phi_j\rangle$. Explicit form of this inner product has been given as Eq.~(B2) in \cite{Yang.17b} (see also Appendix~\ref{appxA}). Using the asymptotic behavior of the modified Bessel function at $x\rightarrow\infty$,
\begin{equation}
 \text{I}_\nu(x)\sim\sqrt{\frac{1}{2\pi x}}e^x,
 \end{equation}
defining $\bm{Q}_{ij}=\bm{R}_i+\bm{R}_j-2\bm{R}_c$ and
\begin{equation}
z_{tij}^{(0)}=\frac{e^2}{4\kappa\sqrt{\pi}a^2_B}\exp\left(-\frac{1}{4 a^2_B}|\bm{R}_i-\bm{R}_j|^2\right),
 \end{equation}
we have (for $R_c\gg a$)
\begin{equation}
\begin{aligned}
z_{tij}=&z_{tij}^{(0)}\exp\left(-\frac{1}{8a^2_B}Q_{ij}^2\right)\text{I}_0\left(\frac{1}{8a^2_B}Q_{ij}^2\right)\\
\sim&\frac{2a_B}{\sqrt{\pi}}z_{tij}^{(0)}\frac{1}{Q_{ij}}.
\end{aligned}\label{eq:smallztij}
\end{equation}
Therefore,
\begin{subequations}
\begin{align}
z_{t12}&\approx\frac{e^2}{4\kappa\pi a_B}\frac{g}{R_c}, \label{eq:zt12}\\
z_{t11}&\approx\frac{e^2}{4\kappa\pi a_B}\frac{1}{\sqrt{R_c^2+a^2+2aR_c\text{cos}\theta}},\label{eq:zt11}\\
z_{t22}&\approx\frac{e^2}{4\kappa\pi a_B}\frac{1}{\sqrt{R_c^2+a^2-2aR_c\text{cos}\theta}}.
\label{eq:zt22}
\end{align}
\end{subequations}

Plugging Eqs.~\eqref{eq:zt12}-\eqref{eq:zt22} into Eq.~\eqref{eq:bigZt12} and expand $\sqrt{R_c^2+a^2\pm2aR_c\text{cos}\theta}$ in the limit of $R_c\gg a$ to the second order, we have
\begin{equation}
Z_{t12}\approx\frac{e^2g}{8\kappa\pi a_B(g^2-1)}\frac{a^2}{R_c^3}\left(3\cos^2\theta-1\right).
\label{eq:Zt12final}
\end{equation}
It is clear that at $\theta=\pm\theta_m=\pm\arccos\left(1/\sqrt{3}\right)$, $Z_{t12}$ approximately vanish and $\delta t\approx0$.

\begin{figure}[t]
(a)\centering\includegraphics[width=0.75\columnwidth]{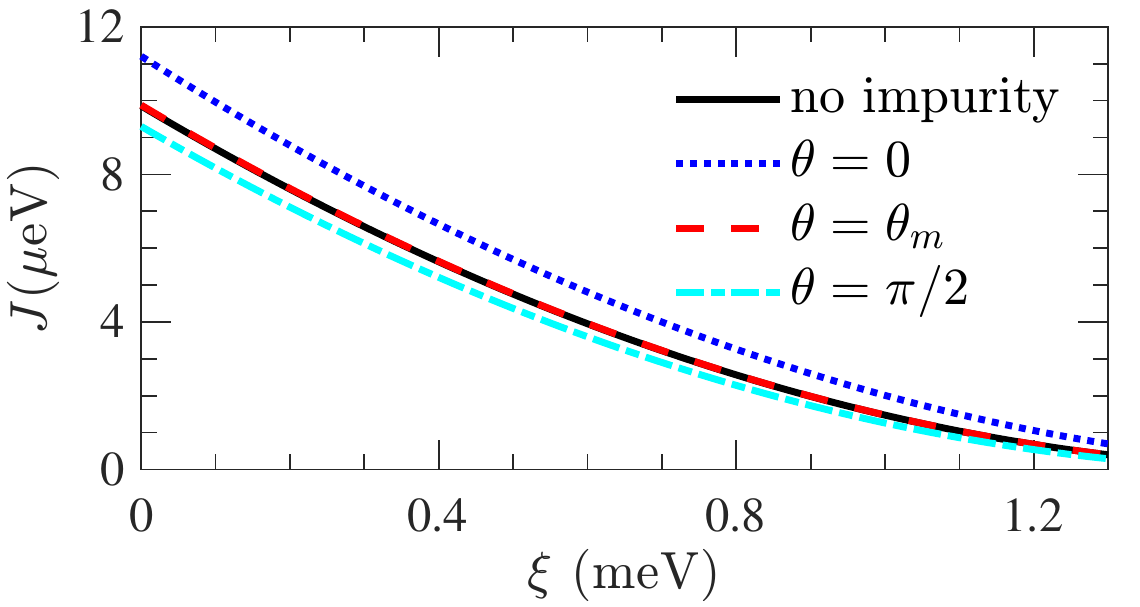}
(b)\centering\includegraphics[width=0.75\columnwidth]{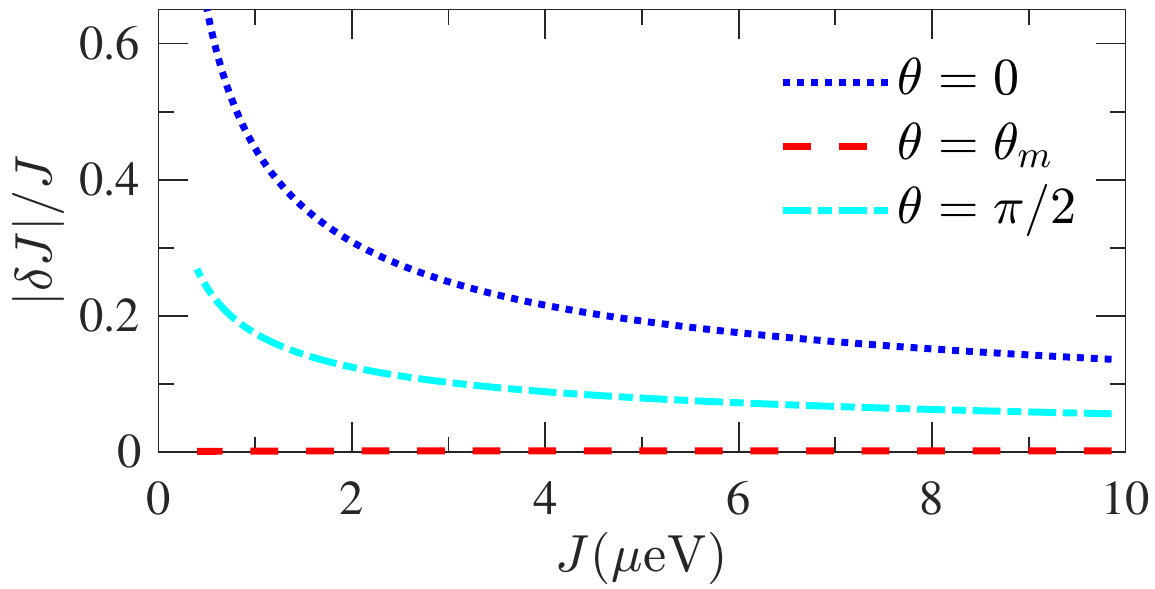}
(c)\centering\includegraphics[width=0.75\columnwidth]{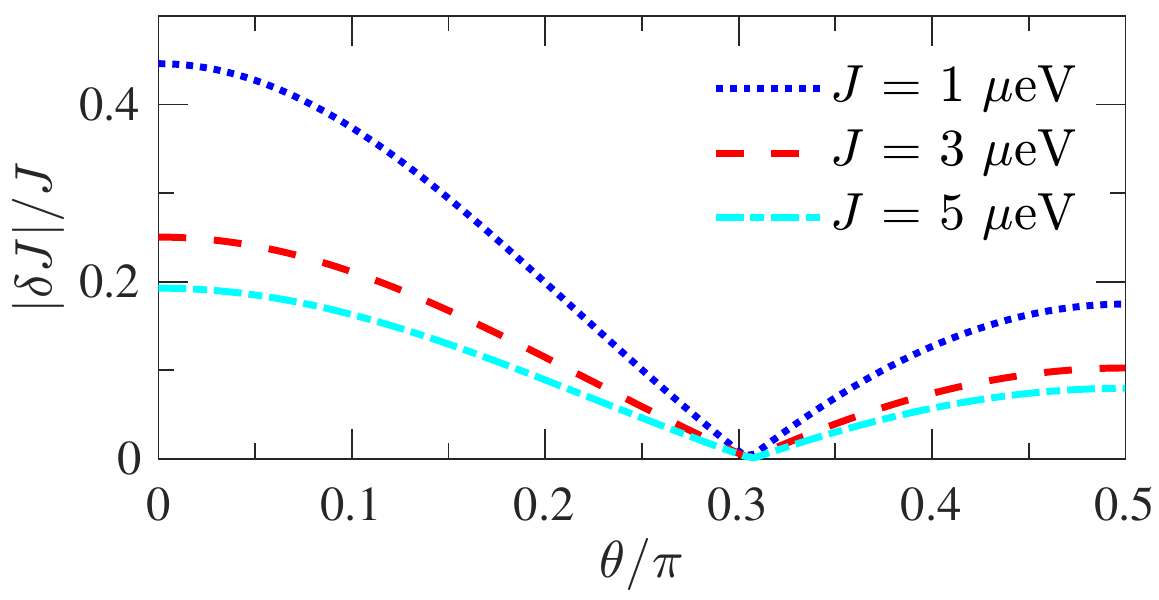}
\caption{(a) The calculated exchange interaction $J$ v.s.~the barrier height $\xi$. The black solid line shows the case without any impurity. Other lines show cases with an impurity at a distance of $R_c = 6a$ from the center of the double dots but along different directions as indicated.
(b) The absolute value of the relative charge noise $|\delta J|/J$ v.s.~$J$ for cases with an impurity. (c) $|\delta J|/J$ v.s. the angle $\theta$ between the impurity vector $\bm{R}_c$ and the polar axis of the double dots, for three different values of $J$ as indicated. Parameters are $a = 100$ nm, $\hbar\omega_0=100$ $\mu$eV.}
\label{fig:oneimpurity}
\end{figure}

Figure~\ref{fig:oneimpurity} shows the exchange interaction and the charge noise caused by the impurity calculated following the method established in \cite{Yang.17b}. From Fig.~\ref{fig:oneimpurity}(a) it is obvious that $J$ decreases as the barrier is raised (increasing $\xi$) without any impurity. An impurity adds charge noise to the exchange interaction and causes shifts in it, the precise value of which depends on the angle between the impurity vector and the polar axis of the double dots, $\theta$. For the three angles presented, the $\theta=0$ case has the greatest noise,  the $\theta=\pi/2$ case carries a smaller shift, but there is hardly any shift for the case $\theta=\theta_m$.  This can be seen more clearly in Fig.~\ref{fig:oneimpurity}(b) which plots $|\delta J|/J$ v.s.~$J$.  $\delta J/J$ is almost zero in the entire range shown for $\theta=\theta_m$ (the actual numerical value is of the order of $10^{-3}$). In Fig.~\ref{fig:oneimpurity}(c) we plot $|\delta J|/J$ v.s.~$\theta$ for three different $J$ values. All curves come down to zero at a little past $0.3\pi$, which is precisely the value of the magic angle ($\approx0.304\pi$).

The result of Eq.~\eqref{eq:Zt12final} can be easily extended to multiple impurities. For $N$ impurities with the same charge $-e$, the total correction to the tunnel coupling $t$ is
\begin{equation}
\sum_{k=1}^NZ_{t12}^{(k)}\approx\frac{e^2ga^2}{8\kappa\pi a_B(g^2-1)}\sum_{k=1}^N\frac{3\cos^2\theta_k-1}{\left(R_c^{(k)}\right)^3},
\label{eq:Zt12multi0}
\end{equation}
where $R_c^{(k)}$ is the length of $\bm{R}_c^{(k)}$, the impurity vector for the $k$th impurity, and $\theta_k$ is the angle between the polar axis of the double dots (here equivalent to $\hat{x}$) and $\bm{R}_c^{(k)}$. 
When $R_c^{(k)}=R_c$ for all $k$ between 1 and $N$, it is clear that the change to the tunnel coupling and subsequently the exchange interaction vanishes provided $\sum_k\cos^2\theta_k=N/3$.

Our finding that the exchange interaction of a barrier-controlled singlet-triplet qubit is insensitive to an impurity positioned at certain directions has interesting implications on a scaled-up qubit array. Typically one fabricate an array of double quantum dots linearly, with the polar axes of all qubits lying along the same line (e.g. the $x$ axis) \cite{Zajac.16,Jones.16}. Our result suggests that in this case or if the qubits are slanted randomly,  the exchange energy of a qubit is affected by its neighboring ones. Nevertheless, the qubits will not affect each other provided if all of them are rotated from $\hat{x}$ by an angle of $\pm\theta_m$,  are being controlled using the barrier method, and are at reasonable distances away from each other ($\gg a$) so that the asymptotic arguments in the aforementioned derivations apply.

\begin{figure}[b]
(a)\centering\includegraphics[width=0.75\columnwidth]{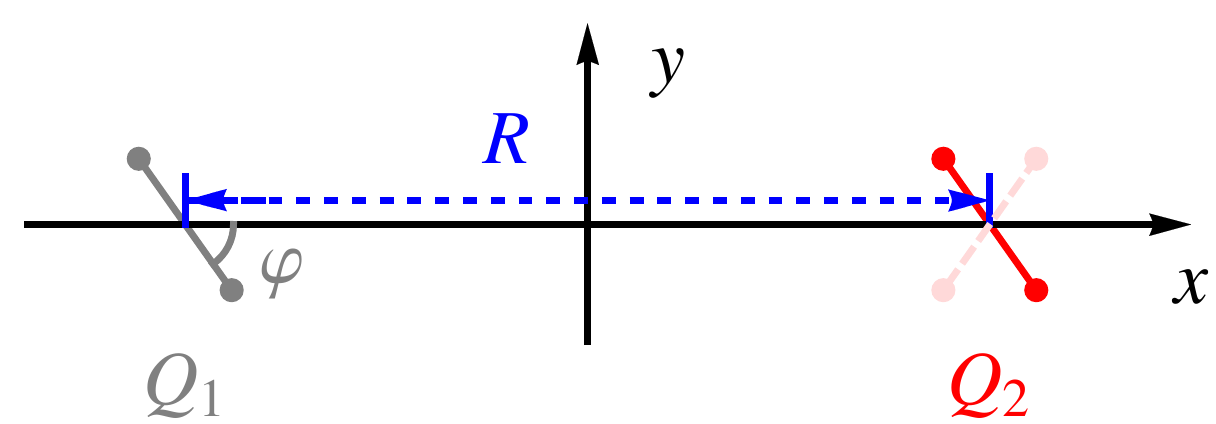}
(b)\centering\includegraphics[width=0.7\columnwidth]{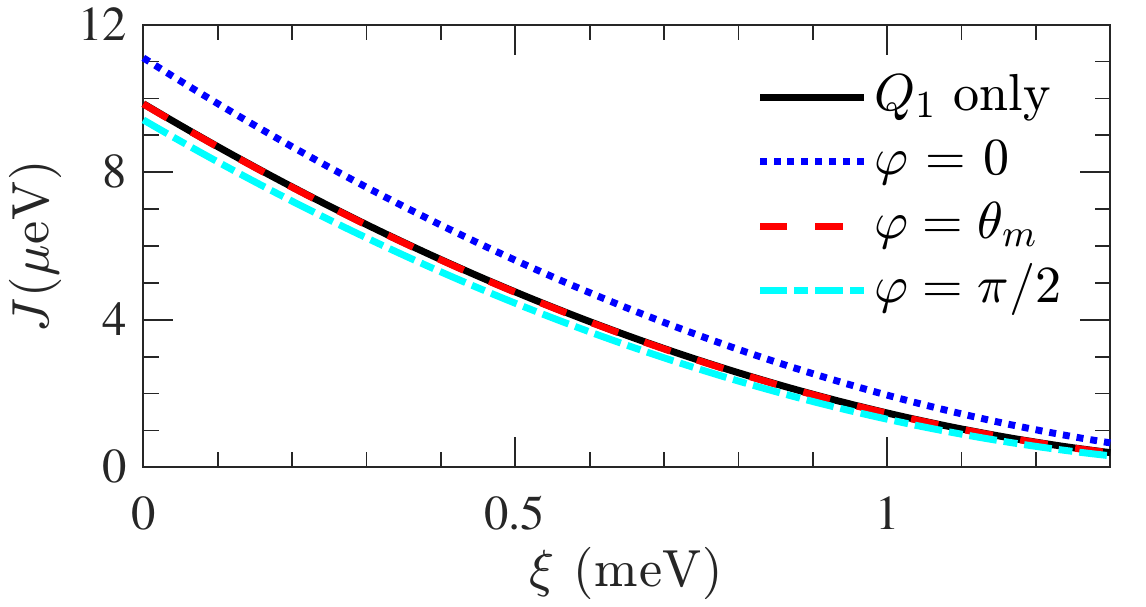}
(c)\centering\includegraphics[width=0.7\columnwidth]{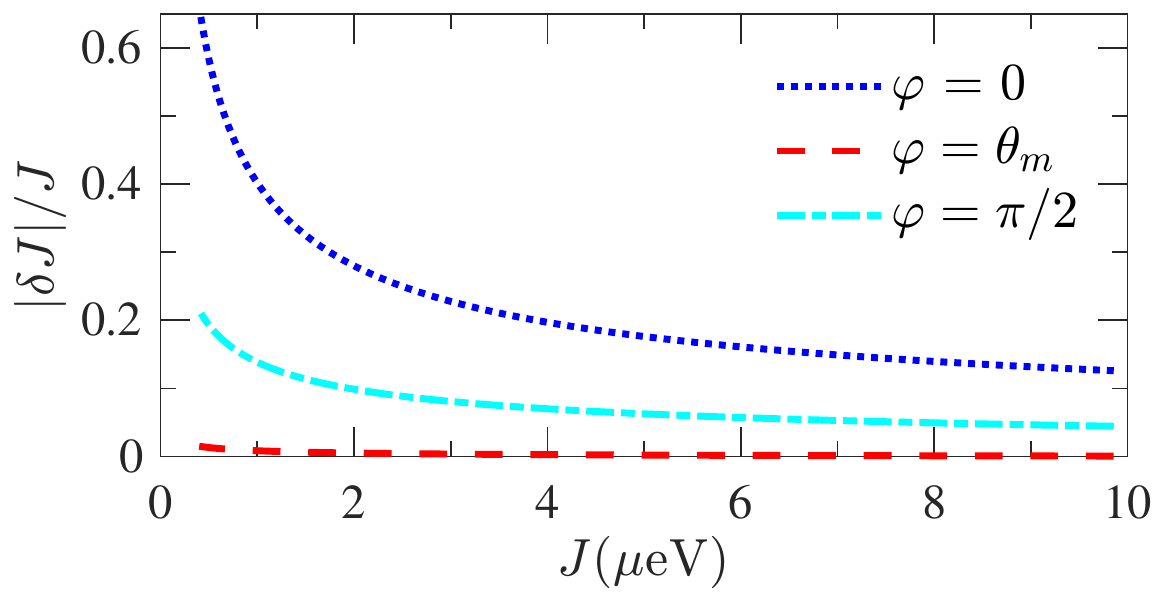}
\caption{(a) Schematics of two pairs of double quantum dots, denoted by $Q_1$ (gray) and $Q_2$ (red), a distance of $R$ apart from each other, both rotated from the $x$ axis by an angle $\varphi$. The light pink dashed line shows an alternative configuration of $Q_2$ (rotated by $-\varphi$).
(b) The calculated exchange interaction $J$ for $Q_1$ v.s.~its barrier height $\xi$. The black solid line shows the case without any other qubit (``$Q_1$ only''). Other lines show cases in presence of another qubit $Q_2$ when both qubits are rotated by an angle $\varphi$ from the $x$ axis.
(c) The absolute value of the relative charge noise $|\delta J|/J$ v.s.~$\varphi$ for cases in presence of $Q_2$. Parameters are $a$ = 100 nm, $\hbar\omega_0=100$ $\mu$eV, $R = 8a$.}
\label{fig:array}
\end{figure}

Figure~\ref{fig:array} presents a numerical demonstration of this argument. Figure~\ref{fig:array}(a) shows a schematic diagram of two double-dot qubits, $Q_1$ and $Q_2$ at a distance of $R$ apart from each other, both rotated from the $x$ axis by an angle $\varphi$. The calculated exchange interaction $J$ for $Q_1$ v.s.~its barrier height $\xi$ is shown in Fig.~\ref{fig:array}(b). The solid black line shows the case $Q_1$ being alone without any other qubit being present, while the dotted, dashed  and dash-dotted lines show cases in presence of $Q_2$ when both qubits are rotated by $\varphi=0$, $\theta_m$, and $\pi/2$. The behavior is rather similar to Fig.~\ref{fig:oneimpurity}(b) that for $\varphi=\theta_m$, almost no drift in $J$ can be seen. The results for $|\delta J|/J$ v.s.~$J$ are plotted in Fig.~\ref{fig:oneimpurity}(c), and the dashed line shows the result of the $\varphi=\theta_m$ case which is close to zero. The actual numerical value is up to 0.015 (for the smallest $J$ concerned) but is below $10^{-3}$ for larger $J$ values. It is clear that the magic angle plays a key role in reducing the cross-talk between qubits. 
We emphasize that for our arguments to apply the qubits do not have to be parallel: the angle between the polar axis of the double dots and $\hat{x}$ can be either $\theta_m$ or $-\theta_m$ for any given qubit, as indicated by the light pink dashed line in Fig.~\ref{fig:array}(a). We have also considered the situation for $R<8a$ and found no significant deviation from our main conclusion as long as $R\gtrsim6a$. The results are presented in Appendix~\ref{appxB}. We note  for a typical four-dot device with all dots equally spaced \cite{Shulman.12}, $R=4a$. In this case, while we are unable to completely suppress the inter-qubit coupling, it can be reduced by about two orders of magnitude [cf. Fig.~\ref{sfig:sf1}(a)] as compared to the traditional lateral design if all qubit are rotated by the magic angle.

During the execution of a typical quantum algorithm on the quantum-dot chain, both single and two qubit manipulations are required. Our results from this simplified model imply that one may switch between the two schemes efficiently by alternating between the barrier and tilt control schemes. When the two-qubit gates are desired, one uses the tilt control method to maximize the capacitive coupling between qubits. When a single-qubit gate is performed on one qubit, it can be made insensitive to the actions of all other qubits provided that the said qubit is operated in the barrier-control manner, and is rotated by the magic angle with respect to the shared reference line. The maximal insensitivity depends on the distance between the qubits but we have shown that in the usual equidistance case ($R=4a$) two orders of magnitude in reduction can happen. We note that our results are derived from a rather simplified model which is still far from an accurate description of an experimental device. For example, the additional Coulomb interaction caused by an impurity may not be of the exact $1/R$ form due to screening effects, and the undesired coupling between qubits is of dipole character, which becomes more significant as the qubits come close. In the situation where coupling between qubits are implemented by floating gates \cite{Trifunovic.12}, the qubit layout must be carefully designed and optimized. Nonetheless, our results show that it is possible, albeit in a rather simplified situation, that unwanted coupling can be at least reduced by judiciously designing the qubit layout.

Finally we note that our arguments not only apply to the singlet-triplet qubit but can also be extended to other quantum-dot qubits, as long as the Rabi frequency of certain rotation around the Bloch sphere is solely dependent on the tunnel coupling $t$. An example is the double-quantum-dot charge qubit \cite{Gorman.05,Oosterkamp.98,Cao.16}: while $z$-axis rotations should be achieved by detuning, the $x$-rotations can be performed at the avoid crossing of two energy levels at zero detuning. The control over the $x$-rotation is therefore insensitive to an impurity along a direction that is $\theta_m$ relative to the polar axis of the double dots.

\section{Conclusion}

In conclusion, we have shown that for a singlet-triplet qubit confined in double quantum dots, its exchange interaction is insensitive to a charged impurity, as long as the impurity vector and the polar axis of the double dots differ by the magic angle, $\theta_m=\arccos\left(1/\sqrt{3}\right)\approx 54.7^\circ$ and the exchange interaction is controlled by the barrier method. 
While our results are derived from a rather simplified model, they show that it is possible to at least reduce the unwanted qubit couplings by carefully designing the qubit layout.

\section*{Acknowledgements}

X.W. thanks J.P. Kestner, E. Barnes and R.E. Throckmorton for discussions, and in particular the hospitality of the Condensed Matter Theory Center at the University of Maryland, College Park, where key advance of this research had taken place.
This work is supported by the 
Research Grants Council of the Hong Kong Special Administrative Region, China (No.~CityU 21300116), the National Natural Science Foundation of China (No.~11604277), and the Guangdong Innovative and Entrepreneurial Research Team Program (No.~2016ZT06D348)

\appendix
\setcounter{equation}{0}

\section{The configuration interaction calculation}\label{appxA}

In this Appendix, we provide more details of our configuration interaction calculation, including the form of the confinement potential.

In this work we follow \cite{Yang.17b} to perform the configuration interaction calculation. We outline the key steps in this section. The full Hamiltonian $H$ includes a single-electron part $H_s$, the interaction between quantum-dot electrons $H_I$, and the interaction between electrons in the quantum dots and the impurity $H_c$:
\begin{equation}
H=H_s+H_I+H_c.
\label{eq:fullH}
\end{equation}
Here, $H_s=h(\bm{r}_1)+h(\bm{r}_2)$,
\begin{equation}
h(\bm{r})=\frac{1}{2m^*}\left[\bm{p}-e\bm{A}(\bm{r})\right]^2+V(\bm{r}).
\label{eq:sH}
\end{equation}
$m^*$ is the effective electron mass (taken to be 
 $0.067m_e$ for GaAs, where $m_e$ is the electron mass) and $\bm{A}$ is the vector potential of the applied magnetic field.
\begin{equation}
H_I=\frac{e^2}{4\pi\kappa|\bm{r}_1-\bm{r}_2|},
\label{eq:Hc}
\end{equation}
where $\kappa=\epsilon\epsilon_0$, $\epsilon=13.1$, and $\epsilon_0$ is the vacuum permittivity.
For an impurity with charge $-e$ at $\bm{R}_c$,
\begin{equation}
H_c=\sum_{i=1}^2\frac{e^2}{4\pi\kappa|\bm{r}_i-\bm{R}_c|}.
\label{eq:im}
\end{equation}

The confinement potential is defined as
\begin{widetext}
\begin{eqnarray}
V(x,y)=
\begin{cases}
-\mu_1+\frac{m^*\omega_0^2}{2}[(x+a)^2+y^2]+\frac{4C+4\mu_1-a^2m^*\omega_0^2}{a^3}(x+a)^3+\frac{-6C-6\mu_1+a^2m^*\omega_0^2}{2a^4}(x+a)^4+G(x,y), \ x\ge0,\cr -\mu_2+\frac{m^*\omega_0^2}{2}[(x-a)^2+y^2]-\frac{4C+4\mu_2-a^2m^*\omega_0^2}{a^3}(x-a)^3+\frac{-6C-6\mu_2+a^2m^*\omega_0^2}{2a^4}(x-a)^4+G(x,y), \ x<0,\end{cases}
\label{eq:potential}
\end{eqnarray}
\end{widetext}
where $C=a^2m^*\omega_0/12$ and $G(x,y)=\xi\exp\left[-8(x^2+y^2)/a^2\right]$. This complicated form is  designed to guarantee that (i) the location of the two wells and the central barrier will not change upon either tilt or barrier control and (ii) the two wells are well approximated by the harmonic oscillator potential
\begin{equation}
V(x,y)|_{(x,y)\rightarrow (\pm a,0)}\approx\frac{m^*\omega_0^2}{2}[(x\pm a)^2+y^2]-\mu_{1,2},
\label{eq:appr}
\end{equation}
with the confinement energy $\omega_0$ fixed throughout the control process. The height of the central potential barrier is solely controlled by $\xi$.

Under the Hund-Mulliken approximation, only the ground states of a harmonic oscillator is considered:
\begin{equation}
\begin{aligned}
\phi_i(\bm{r})=\frac{1}{a_B^{}\sqrt{\pi}}\exp\left[{-\frac{1}{2a_B^2}\left|\bm{r}-\bm{R}_i\right|^2}\right],\quad i=1,2,
\end{aligned}
\label{eq:ground}
\end{equation}
where $a_B\equiv\sqrt{\hbar/(m^*\omega_0)}$ is Fock-Darwin radius. The single-electron wave functions in the double-quantum-dot system are given by:
\begin{equation}
\begin{aligned}
\left\{\psi_1,\text{ }\psi_2\right\}^\text{T}=\mathcal{O}^{-1/2}\left\{\phi_1,\text{ }\phi_2\right\}^\text{T},
\end{aligned}
\label{eq:trans}
\end{equation}
where $\mathcal{O}$ is the overlap matrix: $\mathcal{O}_{l,l^\prime}\equiv\langle\phi_l|\phi_{l^\prime}\rangle$. $\mathcal{O}^{-1/2}$ can be found following \cite{Yang.17a}.

The full Hamiltonian can then be written in the matrix form as
\begin{widetext}
\begin{equation}
H=\left(\begin{array}{cccc}
U_2-2\mu_2+2{Z_t}_2&-t+{Z_t}_{12}&-t+{Z_t}_{12}&0\\
-t+{Z_t}_{12}&U_{12}-\mu_1-\mu_2+{Z_t}_1+{Z_t}_2&0&-t+{Z_t}_{12}\\
-t+{Z_t}_{12}&0&U_{12}-\mu_1-\mu_2+{Z_t}_1+{Z_t}_2&-t+{Z_t}_{12}\\
0&-t+{Z_t}_{12}&-t+{Z_t}_{12}&U_1-2\mu_1+2{Z_t}_1\\
\end{array}
\right),
\label{eq:aHam2}
\end{equation}
\end{widetext}
where $\mu_{1,2}$, $U_{1,2}$, $U_{12}$ and $t$ are Hubbard parameters as in Eq.~\eqref{eq:firstHam},  the $Z_t$ terms are corrections due to the impurity. They are all calculated by taking appropriate inner products. 

In calculating $Z_{t12}$, the following integral is useful:
\begin{widetext}
\begin{equation}
\int{\phi_i(\rr)^*\frac{e^2}{4\pi\kappa|\rr-\bm{R}_C|}\phi_j(\rr)d\rr}=\frac{e^2}{4\kappa\sqrt{\pi}a^2_B}e^{-\frac{1}{4 a^2_B}|\bm{R}_i-\bm{R}_j|^2-\frac{1}{8a^2_B}|\bm{R}_i+\bm{R}_j-2\bm{R}_C|^2}\text{I}_0\left(\frac{1}{8a^2_B}|\bm{R}_i+\bm{R}_j-2\bm{R}_C|^2\right)
\label{eq:AppBimp}
\end{equation}
\end{widetext}
where $\text{I}_0$ is the zeroth-order modified Bessel function of the first kind. This equality has been used to derive Eq.~\eqref{eq:smallztij} in the main text.

\section{Supplemental results for two singlet-triplet qubits}\label{appxB}

Here we discuss, for an array of qubits, how the exchange interaction of one qubit is affected by another when they are closer than the $R=8a$ as discussed in the main text.

In the main text we have considered two pairs of singlet-triplet qubits at a distance of $R$ apart. In Fig.~\ref{fig:array} we have shown results for $R=8a$. Since many of our arguments works in the large $R$ limit, it is an interesting question how our main conclusions may change if $R<8a$.

For singlet-triplet qubits embedded in a typical linear chain with equal spacing between dots, $R=4a$. We therefore calculate $\delta J/J$ for a range of $R$ starting with $R=4a$. The results of $|\delta J|/J$ v.s.~$J$ are shown in Supplementary Fig.~\ref{sfig:sf1}. For all results shown, $|\delta J|/J$ for $\varphi=\theta_m$ is the smallest compared to other ones. 
For $R=4a$, the system is not quite in the asymptotic regime so that $|\delta J|/J$ for $\varphi=\theta_m$ is on the order of 0.1. When $R$ is increased to $5a$, $|\delta J|/J$ for $\varphi=\theta_m$ is already mostly below 0.1 (except for the smallest $J$ value concerned). Further increasing $R$ leads to decrease of the $|\delta J|/J$. In Supplementary Fig.~\ref{sfig:sf1}(c) and (d), $|\delta J|/J$ at $\varphi=\theta_m$ is close to or smaller than 0.01 for reasonably large $J$ used in the quantum computation, suggesting that for $R\gtrsim6a$ the inter-qubit cross-talk is sufficiently small for the purpose of qubit control.

We also plot $|\delta J|/J$ as functions of $R/a$ in  Supplementary Fig.~\ref{sfig:sf2}. Increasing $J$ or $R$ both reduces $|\delta J|/J$, as expected. We can also see that in all cases, $|\delta J|/J$ is very small for $R\gtrsim6a$, suggesting the validity of our arguments presented in the main text. Even for  $4a\le R<6a$, the cross-talk between qubits is already much smaller for $\varphi=\theta_m$ than other angles, so that our results are still useful in cases where qubits must be fabricated close to each other.

\begin{figure}
\centering
\includegraphics[width=0.95\columnwidth]{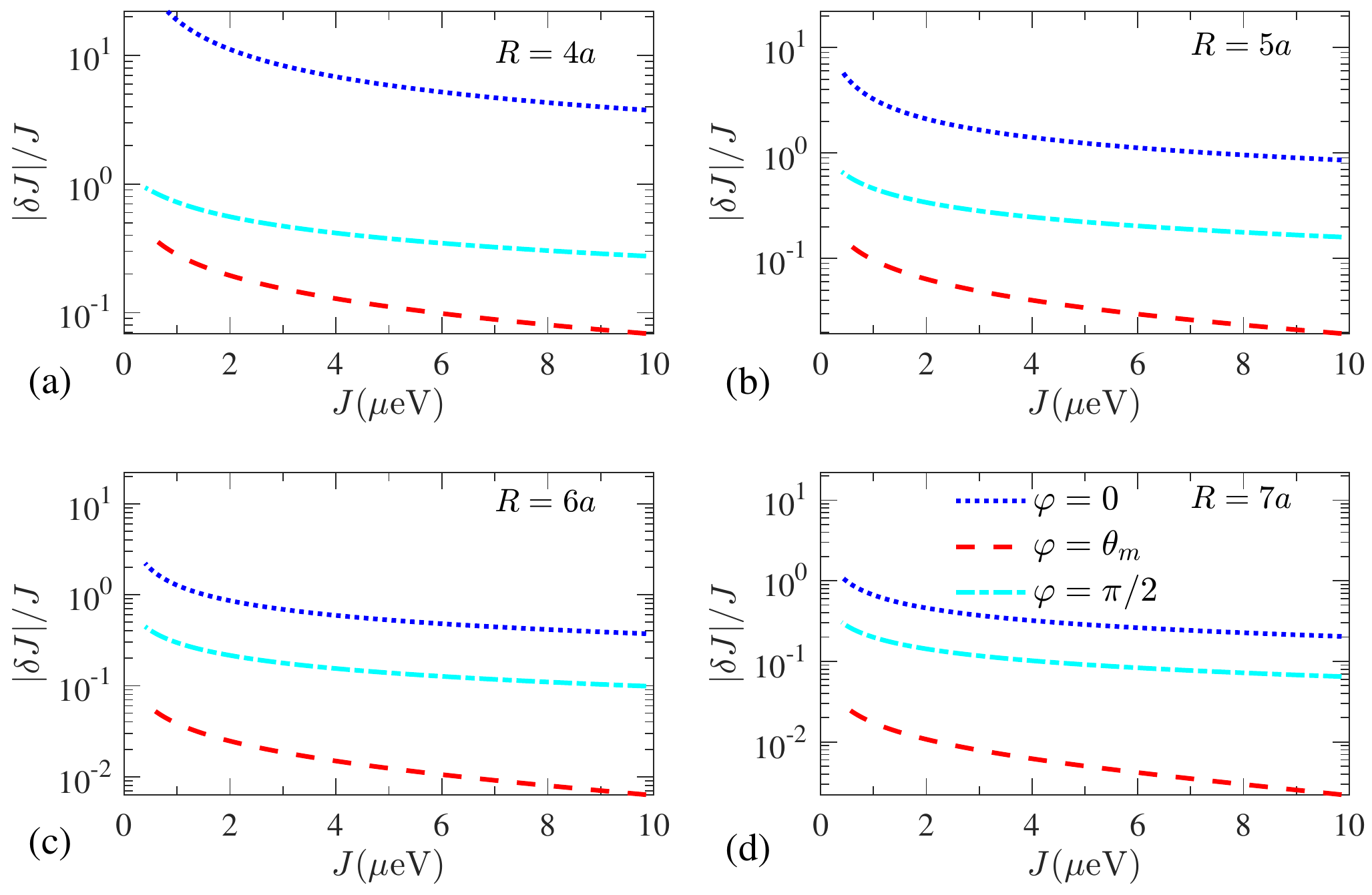}
	\caption{$|\delta J|/J$ v.s.~$J$ for a pair of singlet-triplet qubits at a distance of $R$ apart from each other, the polar axes of which are rotated by $\varphi$ from a shared reference line [cf. Fig.~\ref{fig:array}(a)]. The calculated exchange interaction refers to one of the two qubits. Note different scales of the $y$-axes. (a) $R=4a$. (b) $R=5a$. (c) $R=6a$. (d) $R=7a$.}
\label{sfig:sf1}
\end{figure}

\begin{figure}
\centering
\includegraphics[width=0.95\columnwidth]{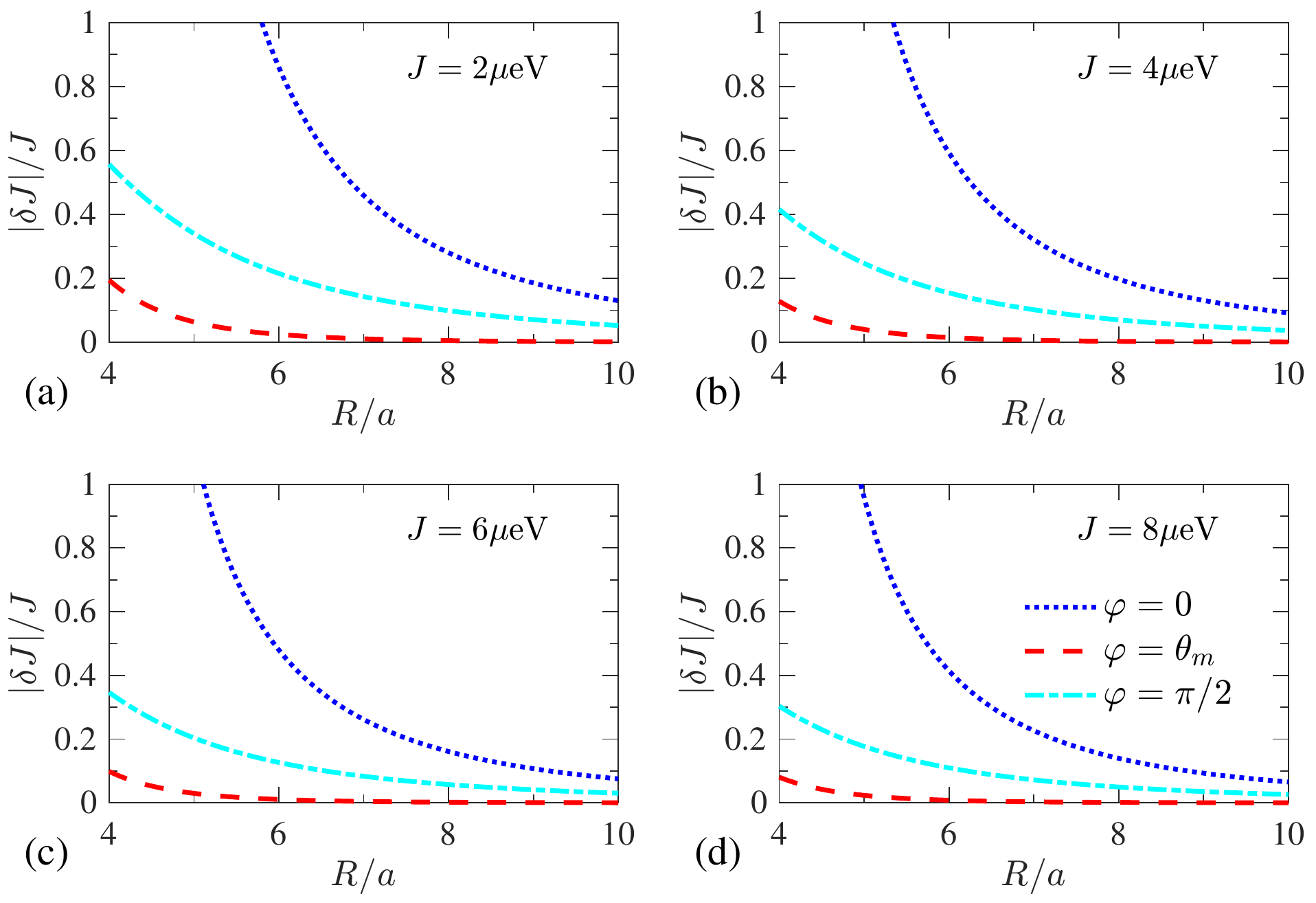}
	\caption{$|\delta J|/J$ v.s.~$R/a$ for a pair of singlet-triplet qubits at a distance of $R$ apart from each other, the polar axes of which are rotated by $\varphi$ from a shared reference line [cf. Fig.~\ref{fig:array}(a)]. The calculated exchange interaction refers to one of the two qubits. (a) $J=2\mu$eV. (b) $J=4\mu$eV. (c) $J=6\mu$eV. (d) $J=8\mu$eV.}
\label{sfig:sf2}
\end{figure}


\begin{thebibliography}{59}%
\makeatletter
\providecommand \@ifxundefined [1]{%
 \@ifx{#1\undefined}
}%
\providecommand \@ifnum [1]{%
 \ifnum #1\expandafter \@firstoftwo
 \else \expandafter \@secondoftwo
 \fi
}%
\providecommand \@ifx [1]{%
 \ifx #1\expandafter \@firstoftwo
 \else \expandafter \@secondoftwo
 \fi
}%
\providecommand \natexlab [1]{#1}%
\providecommand \enquote  [1]{``#1''}%
\providecommand \bibnamefont  [1]{#1}%
\providecommand \bibfnamefont [1]{#1}%
\providecommand \citenamefont [1]{#1}%
\providecommand \href@noop [0]{\@secondoftwo}%
\providecommand \href [0]{\begingroup \@sanitize@url \@href}%
\providecommand \@href[1]{\@@startlink{#1}\@@href}%
\providecommand \@@href[1]{\endgroup#1\@@endlink}%
\providecommand \@sanitize@url [0]{\catcode `\\12\catcode `\$12\catcode
  `\&12\catcode `\#12\catcode `\^12\catcode `\_12\catcode `\%12\relax}%
\providecommand \@@startlink[1]{}%
\providecommand \@@endlink[0]{}%
\providecommand \url  [0]{\begingroup\@sanitize@url \@url }%
\providecommand \@url [1]{\endgroup\@href {#1}{\urlprefix }}%
\providecommand \urlprefix  [0]{URL }%
\providecommand \Eprint [0]{\href }%
\providecommand \doibase [0]{http://dx.doi.org/}%
\providecommand \selectlanguage [0]{\@gobble}%
\providecommand \bibinfo  [0]{\@secondoftwo}%
\providecommand \bibfield  [0]{\@secondoftwo}%
\providecommand \translation [1]{[#1]}%
\providecommand \BibitemOpen [0]{}%
\providecommand \bibitemStop [0]{}%
\providecommand \bibitemNoStop [0]{.\EOS\space}%
\providecommand \EOS [0]{\spacefactor3000\relax}%
\providecommand \BibitemShut  [1]{\csname bibitem#1\endcsname}%
\let\auto@bib@innerbib\@empty
\bibitem [{\citenamefont {van~der Wiel}\ \emph {et~al.}(2002)\citenamefont
  {van~der Wiel}, \citenamefont {De~Franceschi}, \citenamefont {Elzerman},
  \citenamefont {Fujisawa}, \citenamefont {Tarucha},\ and\ \citenamefont
  {Kouwenhoven}}]{vanderWiel.02}%
  \BibitemOpen
  \bibfield  {author} {\bibinfo {author} {\bibfnamefont {W.~G.}\ \bibnamefont
  {van~der Wiel}}, \bibinfo {author} {\bibfnamefont {S.}~\bibnamefont
  {De~Franceschi}}, \bibinfo {author} {\bibfnamefont {J.~M.}\ \bibnamefont
  {Elzerman}}, \bibinfo {author} {\bibfnamefont {T.}~\bibnamefont {Fujisawa}},
  \bibinfo {author} {\bibfnamefont {S.}~\bibnamefont {Tarucha}}, \ and\
  \bibinfo {author} {\bibfnamefont {L.~P.}\ \bibnamefont {Kouwenhoven}},\
  }\href {\doibase 10.1103/RevModPhys.75.1} {\bibfield  {journal} {\bibinfo
  {journal} {Rev. Mod. Phys.}\ }\textbf {\bibinfo {volume} {75}},\ \bibinfo
  {pages} {1} (\bibinfo {year} {2002})}\BibitemShut {NoStop}%
\bibitem [{\citenamefont {Waugh}\ \emph {et~al.}(1995)\citenamefont {Waugh},
  \citenamefont {Berry}, \citenamefont {Mar}, \citenamefont {Westervelt},
  \citenamefont {Campman},\ and\ \citenamefont {Gossard}}]{Waugh.95}%
  \BibitemOpen
  \bibfield  {author} {\bibinfo {author} {\bibfnamefont {F.~R.}\ \bibnamefont
  {Waugh}}, \bibinfo {author} {\bibfnamefont {M.~J.}\ \bibnamefont {Berry}},
  \bibinfo {author} {\bibfnamefont {D.~J.}\ \bibnamefont {Mar}}, \bibinfo
  {author} {\bibfnamefont {R.~M.}\ \bibnamefont {Westervelt}}, \bibinfo
  {author} {\bibfnamefont {K.~L.}\ \bibnamefont {Campman}}, \ and\ \bibinfo
  {author} {\bibfnamefont {A.~C.}\ \bibnamefont {Gossard}},\ }\href
  {https://link.aps.org/doi/10.1103/PhysRevLett.75.705} {\bibfield  {journal}
  {\bibinfo  {journal} {Phys. Rev. Lett.}\ }\textbf {\bibinfo {volume} {75}},\
  \bibinfo {pages} {705} (\bibinfo {year} {1995})}\BibitemShut {NoStop}%
\bibitem [{\citenamefont {Matveev}\ \emph {et~al.}(1996)\citenamefont
  {Matveev}, \citenamefont {Glazman},\ and\ \citenamefont
  {Baranger}}]{Matveev.96}%
  \BibitemOpen
  \bibfield  {author} {\bibinfo {author} {\bibfnamefont {K.~A.}\ \bibnamefont
  {Matveev}}, \bibinfo {author} {\bibfnamefont {L.~I.}\ \bibnamefont
  {Glazman}}, \ and\ \bibinfo {author} {\bibfnamefont {H.~U.}\ \bibnamefont
  {Baranger}},\ }\href {\doibase 10.1103/PhysRevB.54.5637} {\bibfield
  {journal} {\bibinfo  {journal} {Phys. Rev. B}\ }\textbf {\bibinfo {volume}
  {54}},\ \bibinfo {pages} {5637} (\bibinfo {year} {1996})}\BibitemShut
  {NoStop}%
\bibitem [{\citenamefont {Pohjola}\ \emph {et~al.}(2001)\citenamefont
  {Pohjola}, \citenamefont {Schoeller},\ and\ \citenamefont
  {Sch{\"o}n}}]{Pohjola.01}%
  \BibitemOpen
  \bibfield  {author} {\bibinfo {author} {\bibfnamefont {T.}~\bibnamefont
  {Pohjola}}, \bibinfo {author} {\bibfnamefont {H.}~\bibnamefont {Schoeller}},
  \ and\ \bibinfo {author} {\bibfnamefont {G.}~\bibnamefont {Sch{\"o}n}},\
  }\href {\doibase 10.1209/epl/i2001-00301-2} {\bibfield  {journal} {\bibinfo
  {journal} {Europhys. Lett.}\ }\textbf {\bibinfo {volume} {54}},\ \bibinfo
  {pages} {241} (\bibinfo {year} {2001})}\BibitemShut {NoStop}%
\bibitem [{\citenamefont {Nielsen}\ and\ \citenamefont
  {Chuang}(2010)}]{nielsen2010quantum}%
  \BibitemOpen
  \bibfield  {author} {\bibinfo {author} {\bibfnamefont {M.~A.}\ \bibnamefont
  {Nielsen}}\ and\ \bibinfo {author} {\bibfnamefont {I.~L.}\ \bibnamefont
  {Chuang}},\ }\href@noop {} {\emph {\bibinfo {title} {Quantum computation and
  quantum information}}}\ (\bibinfo  {publisher} {Cambridge university press},\
  \bibinfo {year} {2010})\BibitemShut {NoStop}%
\bibitem [{\citenamefont {Petta}\ \emph {et~al.}(2005)\citenamefont {Petta},
  \citenamefont {Johnson}, \citenamefont {Taylor}, \citenamefont {Laird},
  \citenamefont {Yacoby}, \citenamefont {Lukin}, \citenamefont {Marcus},
  \citenamefont {Hanson},\ and\ \citenamefont {Gossard}}]{Petta.05}%
  \BibitemOpen
  \bibfield  {author} {\bibinfo {author} {\bibfnamefont {J.~R.}\ \bibnamefont
  {Petta}}, \bibinfo {author} {\bibfnamefont {A.~C.}\ \bibnamefont {Johnson}},
  \bibinfo {author} {\bibfnamefont {J.~M.}\ \bibnamefont {Taylor}}, \bibinfo
  {author} {\bibfnamefont {E.~A.}\ \bibnamefont {Laird}}, \bibinfo {author}
  {\bibfnamefont {A.}~\bibnamefont {Yacoby}}, \bibinfo {author} {\bibfnamefont
  {M.~D.}\ \bibnamefont {Lukin}}, \bibinfo {author} {\bibfnamefont {C.~M.}\
  \bibnamefont {Marcus}}, \bibinfo {author} {\bibfnamefont {M.~P.}\
  \bibnamefont {Hanson}}, \ and\ \bibinfo {author} {\bibfnamefont {A.~C.}\
  \bibnamefont {Gossard}},\ }\href {\doibase 10.1126/science.1116955}
  {\bibfield  {journal} {\bibinfo  {journal} {Science}\ }\textbf {\bibinfo
  {volume} {309}},\ \bibinfo {pages} {2180} (\bibinfo {year}
  {2005})}\BibitemShut {NoStop}%
\bibitem [{\citenamefont {Bluhm}\ \emph
  {et~al.}(2010{\natexlab{a}})\citenamefont {Bluhm}, \citenamefont {Foletti},
  \citenamefont {Neder}, \citenamefont {Rudner}, \citenamefont {Mahalu},
  \citenamefont {Umansky},\ and\ \citenamefont {Yacoby}}]{Bluhm.10b}%
  \BibitemOpen
  \bibfield  {author} {\bibinfo {author} {\bibfnamefont {H.}~\bibnamefont
  {Bluhm}}, \bibinfo {author} {\bibfnamefont {S.}~\bibnamefont {Foletti}},
  \bibinfo {author} {\bibfnamefont {I.}~\bibnamefont {Neder}}, \bibinfo
  {author} {\bibfnamefont {M.}~\bibnamefont {Rudner}}, \bibinfo {author}
  {\bibfnamefont {D.}~\bibnamefont {Mahalu}}, \bibinfo {author} {\bibfnamefont
  {V.}~\bibnamefont {Umansky}}, \ and\ \bibinfo {author} {\bibfnamefont
  {A.}~\bibnamefont {Yacoby}},\ }\href {\doibase 10.1038/NPHYS1856} {\bibfield
  {journal} {\bibinfo  {journal} {Nature Phys.}\ }\textbf {\bibinfo {volume}
  {7}},\ \bibinfo {pages} {109} (\bibinfo {year}
  {2010}{\natexlab{a}})}\BibitemShut {NoStop}%
\bibitem [{\citenamefont {Barthel}\ \emph {et~al.}(2010)\citenamefont
  {Barthel}, \citenamefont {Medford}, \citenamefont {Marcus}, \citenamefont
  {Hanson},\ and\ \citenamefont {Gossard}}]{Barthel.10}%
  \BibitemOpen
  \bibfield  {author} {\bibinfo {author} {\bibfnamefont {C.}~\bibnamefont
  {Barthel}}, \bibinfo {author} {\bibfnamefont {J.}~\bibnamefont {Medford}},
  \bibinfo {author} {\bibfnamefont {C.~M.}\ \bibnamefont {Marcus}}, \bibinfo
  {author} {\bibfnamefont {M.~P.}\ \bibnamefont {Hanson}}, \ and\ \bibinfo
  {author} {\bibfnamefont {A.~C.}\ \bibnamefont {Gossard}},\ }\href
  {https://link.aps.org/doi/10.1103/PhysRevLett.105.266808} {\bibfield
  {journal} {\bibinfo  {journal} {Phys. Rev. Lett.}\ }\textbf {\bibinfo
  {volume} {105}},\ \bibinfo {pages} {266808} (\bibinfo {year}
  {2010})}\BibitemShut {NoStop}%
\bibitem [{\citenamefont {Maune}\ \emph {et~al.}(2012)\citenamefont {Maune},
  \citenamefont {Borselli}, \citenamefont {Huang}, \citenamefont {Ladd},
  \citenamefont {Deelman}, \citenamefont {Holabird}, \citenamefont {Kiselev},
  \citenamefont {Alvarado-Rodriguez}, \citenamefont {Ross}, \citenamefont
  {Schmitz}, \citenamefont {Sokolich}, \citenamefont {Watson}, \citenamefont
  {Gyure},\ and\ \citenamefont {Hunter}}]{Maune.12}%
  \BibitemOpen
  \bibfield  {author} {\bibinfo {author} {\bibfnamefont {B.~M.}\ \bibnamefont
  {Maune}}, \bibinfo {author} {\bibfnamefont {M.~G.}\ \bibnamefont {Borselli}},
  \bibinfo {author} {\bibfnamefont {B.}~\bibnamefont {Huang}}, \bibinfo
  {author} {\bibfnamefont {T.~D.}\ \bibnamefont {Ladd}}, \bibinfo {author}
  {\bibfnamefont {P.~W.}\ \bibnamefont {Deelman}}, \bibinfo {author}
  {\bibfnamefont {K.~S.}\ \bibnamefont {Holabird}}, \bibinfo {author}
  {\bibfnamefont {A.~A.}\ \bibnamefont {Kiselev}}, \bibinfo {author}
  {\bibfnamefont {I.}~\bibnamefont {Alvarado-Rodriguez}}, \bibinfo {author}
  {\bibfnamefont {R.~S.}\ \bibnamefont {Ross}}, \bibinfo {author}
  {\bibfnamefont {A.~E.}\ \bibnamefont {Schmitz}}, \bibinfo {author}
  {\bibfnamefont {M.}~\bibnamefont {Sokolich}}, \bibinfo {author}
  {\bibfnamefont {C.~A.}\ \bibnamefont {Watson}}, \bibinfo {author}
  {\bibfnamefont {M.~F.}\ \bibnamefont {Gyure}}, \ and\ \bibinfo {author}
  {\bibfnamefont {A.~T.}\ \bibnamefont {Hunter}},\ }\href
  {http://dx.doi.org/10.1038/nature10707} {\bibfield  {journal} {\bibinfo
  {journal} {Nature}\ }\textbf {\bibinfo {volume} {481}},\ \bibinfo {pages}
  {344} (\bibinfo {year} {2012})}\BibitemShut {NoStop}%
\bibitem [{\citenamefont {Pla}\ \emph {et~al.}(2013)\citenamefont {Pla},
  \citenamefont {Tan}, \citenamefont {Dehollain}, \citenamefont {Lim},
  \citenamefont {Morton}, \citenamefont {Zwanenburg}, \citenamefont {Jamieson},
  \citenamefont {Dzurak},\ and\ \citenamefont {Morello}}]{Pla.13}%
  \BibitemOpen
  \bibfield  {author} {\bibinfo {author} {\bibfnamefont {J.~J.}\ \bibnamefont
  {Pla}}, \bibinfo {author} {\bibfnamefont {K.~Y.}\ \bibnamefont {Tan}},
  \bibinfo {author} {\bibfnamefont {J.~P.}\ \bibnamefont {Dehollain}}, \bibinfo
  {author} {\bibfnamefont {W.~H.}\ \bibnamefont {Lim}}, \bibinfo {author}
  {\bibfnamefont {J.~J.~L.}\ \bibnamefont {Morton}}, \bibinfo {author}
  {\bibfnamefont {F.~A.}\ \bibnamefont {Zwanenburg}}, \bibinfo {author}
  {\bibfnamefont {D.~N.}\ \bibnamefont {Jamieson}}, \bibinfo {author}
  {\bibfnamefont {A.~S.}\ \bibnamefont {Dzurak}}, \ and\ \bibinfo {author}
  {\bibfnamefont {A.}~\bibnamefont {Morello}},\ }\href
  {http://dx.doi.org/10.1038/nature12011} {\bibfield  {journal} {\bibinfo
  {journal} {Nature}\ }\textbf {\bibinfo {volume} {496}},\ \bibinfo {pages}
  {334} (\bibinfo {year} {2013})}\BibitemShut {NoStop}%
\bibitem [{\citenamefont {Muhonen}\ \emph {et~al.}(2014)\citenamefont
  {Muhonen}, \citenamefont {Dehollain}, \citenamefont {Laucht}, \citenamefont
  {Hudson}, \citenamefont {Kalra}, \citenamefont {Sekiguchi}, \citenamefont
  {Itoh}, \citenamefont {Jamieson}, \citenamefont {McCallum}, \citenamefont
  {Dzurak},\ and\ \citenamefont {Morello}}]{Muhonen.14}%
  \BibitemOpen
  \bibfield  {author} {\bibinfo {author} {\bibfnamefont {J.~T.}\ \bibnamefont
  {Muhonen}}, \bibinfo {author} {\bibfnamefont {J.~P.}\ \bibnamefont
  {Dehollain}}, \bibinfo {author} {\bibfnamefont {A.}~\bibnamefont {Laucht}},
  \bibinfo {author} {\bibfnamefont {F.~E.}\ \bibnamefont {Hudson}}, \bibinfo
  {author} {\bibfnamefont {R.}~\bibnamefont {Kalra}}, \bibinfo {author}
  {\bibfnamefont {T.}~\bibnamefont {Sekiguchi}}, \bibinfo {author}
  {\bibfnamefont {K.~M.}\ \bibnamefont {Itoh}}, \bibinfo {author}
  {\bibfnamefont {D.~N.}\ \bibnamefont {Jamieson}}, \bibinfo {author}
  {\bibfnamefont {J.~C.}\ \bibnamefont {McCallum}}, \bibinfo {author}
  {\bibfnamefont {A.~S.}\ \bibnamefont {Dzurak}}, \ and\ \bibinfo {author}
  {\bibfnamefont {A.}~\bibnamefont {Morello}},\ }\href
  {http://dx.doi.org/10.1038/nnano.2014.211} {\bibfield  {journal} {\bibinfo
  {journal} {Nat. Nanotechnol.}\ }\textbf {\bibinfo {volume} {9}},\ \bibinfo
  {pages} {986} (\bibinfo {year} {2014})}\BibitemShut {NoStop}%
\bibitem [{\citenamefont {Kim}\ \emph {et~al.}(2014)\citenamefont {Kim},
  \citenamefont {Shi}, \citenamefont {Simmons}, \citenamefont {Ward},
  \citenamefont {Prance}, \citenamefont {Koh}, \citenamefont {Gamble},
  \citenamefont {Savage}, \citenamefont {Lagally}, \citenamefont {Friesen},
  \citenamefont {Coppersmith},\ and\ \citenamefont {Eriksson}}]{Kim.14}%
  \BibitemOpen
  \bibfield  {author} {\bibinfo {author} {\bibfnamefont {D.}~\bibnamefont
  {Kim}}, \bibinfo {author} {\bibfnamefont {Z.}~\bibnamefont {Shi}}, \bibinfo
  {author} {\bibfnamefont {C.~B.}\ \bibnamefont {Simmons}}, \bibinfo {author}
  {\bibfnamefont {D.~R.}\ \bibnamefont {Ward}}, \bibinfo {author}
  {\bibfnamefont {J.~R.}\ \bibnamefont {Prance}}, \bibinfo {author}
  {\bibfnamefont {T.~S.}\ \bibnamefont {Koh}}, \bibinfo {author} {\bibfnamefont
  {J.~K.}\ \bibnamefont {Gamble}}, \bibinfo {author} {\bibfnamefont {D.~E.}\
  \bibnamefont {Savage}}, \bibinfo {author} {\bibfnamefont {M.~G.}\
  \bibnamefont {Lagally}}, \bibinfo {author} {\bibfnamefont {M.}~\bibnamefont
  {Friesen}}, \bibinfo {author} {\bibfnamefont {S.~N.}\ \bibnamefont
  {Coppersmith}}, \ and\ \bibinfo {author} {\bibfnamefont {M.~A.}\ \bibnamefont
  {Eriksson}},\ }\href {http://dx.doi.org/10.1038/nature13407} {\bibfield
  {journal} {\bibinfo  {journal} {Nature}\ }\textbf {\bibinfo {volume} {511}},\
  \bibinfo {pages} {70} (\bibinfo {year} {2014})}\BibitemShut {NoStop}%
\bibitem [{\citenamefont {Kawakami}\ \emph {et~al.}(2016)\citenamefont
  {Kawakami}, \citenamefont {Jullien}, \citenamefont {Scarlino}, \citenamefont
  {Ward}, \citenamefont {Savage}, \citenamefont {Lagally}, \citenamefont
  {Dobrovitski}, \citenamefont {Friesen}, \citenamefont {Coppersmith},
  \citenamefont {Eriksson},\ and\ \citenamefont {Vandersypen}}]{Kawakami.16}%
  \BibitemOpen
  \bibfield  {author} {\bibinfo {author} {\bibfnamefont {E.}~\bibnamefont
  {Kawakami}}, \bibinfo {author} {\bibfnamefont {T.}~\bibnamefont {Jullien}},
  \bibinfo {author} {\bibfnamefont {P.}~\bibnamefont {Scarlino}}, \bibinfo
  {author} {\bibfnamefont {D.~R.}\ \bibnamefont {Ward}}, \bibinfo {author}
  {\bibfnamefont {D.~E.}\ \bibnamefont {Savage}}, \bibinfo {author}
  {\bibfnamefont {M.~G.}\ \bibnamefont {Lagally}}, \bibinfo {author}
  {\bibfnamefont {V.~V.}\ \bibnamefont {Dobrovitski}}, \bibinfo {author}
  {\bibfnamefont {M.}~\bibnamefont {Friesen}}, \bibinfo {author} {\bibfnamefont
  {S.~N.}\ \bibnamefont {Coppersmith}}, \bibinfo {author} {\bibfnamefont
  {M.~A.}\ \bibnamefont {Eriksson}}, \ and\ \bibinfo {author} {\bibfnamefont
  {L.~M.~K.}\ \bibnamefont {Vandersypen}},\ }\href
  {http://www.pnas.org/content/113/42/11738.abstract} {\bibfield  {journal}
  {\bibinfo  {journal} {Proc. Natl. Acad. Sci. U.S.A.}\ }\textbf {\bibinfo
  {volume} {113}},\ \bibinfo {pages} {11738} (\bibinfo {year}
  {2016})}\BibitemShut {NoStop}%
\bibitem [{\citenamefont {Taylor}\ \emph {et~al.}(2005)\citenamefont {Taylor},
  \citenamefont {Engel}, \citenamefont {Dur}, \citenamefont {Yacoby},
  \citenamefont {Marcus}, \citenamefont {Zoller},\ and\ \citenamefont
  {Lukin}}]{Taylor.05}%
  \BibitemOpen
  \bibfield  {author} {\bibinfo {author} {\bibfnamefont {J.~M.}\ \bibnamefont
  {Taylor}}, \bibinfo {author} {\bibfnamefont {H.~A.}\ \bibnamefont {Engel}},
  \bibinfo {author} {\bibfnamefont {W.}~\bibnamefont {Dur}}, \bibinfo {author}
  {\bibfnamefont {A.}~\bibnamefont {Yacoby}}, \bibinfo {author} {\bibfnamefont
  {C.~M.}\ \bibnamefont {Marcus}}, \bibinfo {author} {\bibfnamefont
  {P.}~\bibnamefont {Zoller}}, \ and\ \bibinfo {author} {\bibfnamefont {M.~D.}\
  \bibnamefont {Lukin}},\ }\href {\doibase 10.1038/nphys174} {\bibfield
  {journal} {\bibinfo  {journal} {Nat. Phys.}\ }\textbf {\bibinfo {volume}
  {1}},\ \bibinfo {pages} {177} (\bibinfo {year} {2005})}\BibitemShut {NoStop}%
\bibitem [{\citenamefont {Gorman}\ \emph {et~al.}(2005)\citenamefont {Gorman},
  \citenamefont {Hasko},\ and\ \citenamefont {Williams}}]{Gorman.05}%
  \BibitemOpen
  \bibfield  {author} {\bibinfo {author} {\bibfnamefont {J.}~\bibnamefont
  {Gorman}}, \bibinfo {author} {\bibfnamefont {D.~G.}\ \bibnamefont {Hasko}}, \
  and\ \bibinfo {author} {\bibfnamefont {D.~A.}\ \bibnamefont {Williams}},\
  }\href {\doibase 10.1103/PhysRevLett.95.090502} {\bibfield  {journal}
  {\bibinfo  {journal} {Phys. Rev. Lett.}\ }\textbf {\bibinfo {volume} {95}},\
  \bibinfo {pages} {090502} (\bibinfo {year} {2005})}\BibitemShut {NoStop}%
\bibitem [{\citenamefont {Loss}\ and\ \citenamefont
  {DiVincenzo}(1998)}]{Loss.98}%
  \BibitemOpen
  \bibfield  {author} {\bibinfo {author} {\bibfnamefont {D.}~\bibnamefont
  {Loss}}\ and\ \bibinfo {author} {\bibfnamefont {D.~P.}\ \bibnamefont
  {DiVincenzo}},\ }\href {\doibase 10.1103/PhysRevA.57.120} {\bibfield
  {journal} {\bibinfo  {journal} {Phys. Rev. A}\ }\textbf {\bibinfo {volume}
  {57}},\ \bibinfo {pages} {120} (\bibinfo {year} {1998})}\BibitemShut
  {NoStop}%
\bibitem [{\citenamefont {DiVincenzo}\ \emph {et~al.}(2000)\citenamefont
  {DiVincenzo}, \citenamefont {Bacon}, \citenamefont {Kempe}, \citenamefont
  {Burkard},\ and\ \citenamefont {Whaley}}]{DiVincenzo.00}%
  \BibitemOpen
  \bibfield  {author} {\bibinfo {author} {\bibfnamefont {D.~P.}\ \bibnamefont
  {DiVincenzo}}, \bibinfo {author} {\bibfnamefont {D.}~\bibnamefont {Bacon}},
  \bibinfo {author} {\bibfnamefont {J.}~\bibnamefont {Kempe}}, \bibinfo
  {author} {\bibfnamefont {G.}~\bibnamefont {Burkard}}, \ and\ \bibinfo
  {author} {\bibfnamefont {K.~B.}\ \bibnamefont {Whaley}},\ }\href
  {http://dx.doi.org/10.1038/35042541} {\bibfield  {journal} {\bibinfo
  {journal} {Nature}\ }\textbf {\bibinfo {volume} {408}},\ \bibinfo {pages}
  {339} (\bibinfo {year} {2000})}\BibitemShut {NoStop}%
\bibitem [{\citenamefont {Levy}(2002)}]{Levy.02}%
  \BibitemOpen
  \bibfield  {author} {\bibinfo {author} {\bibfnamefont {J.}~\bibnamefont
  {Levy}},\ }\href {\doibase 10.1103/PhysRevLett.89.147902} {\bibfield
  {journal} {\bibinfo  {journal} {Phys. Rev. Lett.}\ }\textbf {\bibinfo
  {volume} {89}},\ \bibinfo {pages} {147902} (\bibinfo {year}
  {2002})}\BibitemShut {NoStop}%
\bibitem [{\citenamefont {Shi}\ \emph {et~al.}(2012)\citenamefont {Shi},
  \citenamefont {Simmons}, \citenamefont {Prance}, \citenamefont {Gamble},
  \citenamefont {Koh}, \citenamefont {Shim}, \citenamefont {Hu}, \citenamefont
  {Savage}, \citenamefont {Lagally}, \citenamefont {Eriksson}, \citenamefont
  {Friesen},\ and\ \citenamefont {Coppersmith}}]{Shi.12}%
  \BibitemOpen
  \bibfield  {author} {\bibinfo {author} {\bibfnamefont {Z.}~\bibnamefont
  {Shi}}, \bibinfo {author} {\bibfnamefont {C.~B.}\ \bibnamefont {Simmons}},
  \bibinfo {author} {\bibfnamefont {J.~R.}\ \bibnamefont {Prance}}, \bibinfo
  {author} {\bibfnamefont {J.~K.}\ \bibnamefont {Gamble}}, \bibinfo {author}
  {\bibfnamefont {T.~S.}\ \bibnamefont {Koh}}, \bibinfo {author} {\bibfnamefont
  {Y.-P.}\ \bibnamefont {Shim}}, \bibinfo {author} {\bibfnamefont
  {X.}~\bibnamefont {Hu}}, \bibinfo {author} {\bibfnamefont {D.~E.}\
  \bibnamefont {Savage}}, \bibinfo {author} {\bibfnamefont {M.~G.}\
  \bibnamefont {Lagally}}, \bibinfo {author} {\bibfnamefont {M.~A.}\
  \bibnamefont {Eriksson}}, \bibinfo {author} {\bibfnamefont {M.}~\bibnamefont
  {Friesen}}, \ and\ \bibinfo {author} {\bibfnamefont {S.~N.}\ \bibnamefont
  {Coppersmith}},\ }\href {\doibase 10.1103/PhysRevLett.108.140503} {\bibfield
  {journal} {\bibinfo  {journal} {Phys. Rev. Lett.}\ }\textbf {\bibinfo
  {volume} {108}},\ \bibinfo {pages} {140503} (\bibinfo {year}
  {2012})}\BibitemShut {NoStop}%
\bibitem [{\citenamefont {Foletti}\ \emph {et~al.}(2009)\citenamefont
  {Foletti}, \citenamefont {Bluhm}, \citenamefont {Mahalu}, \citenamefont
  {Umansky},\ and\ \citenamefont {Yacoby}}]{Foletti.09}%
  \BibitemOpen
  \bibfield  {author} {\bibinfo {author} {\bibfnamefont {S.}~\bibnamefont
  {Foletti}}, \bibinfo {author} {\bibfnamefont {H.}~\bibnamefont {Bluhm}},
  \bibinfo {author} {\bibfnamefont {D.}~\bibnamefont {Mahalu}}, \bibinfo
  {author} {\bibfnamefont {V.}~\bibnamefont {Umansky}}, \ and\ \bibinfo
  {author} {\bibfnamefont {A.}~\bibnamefont {Yacoby}},\ }\href
  {http://dx.doi.org/10.1038/nphys1424} {\bibfield  {journal} {\bibinfo
  {journal} {Nature Phys.}\ }\textbf {\bibinfo {volume} {5}},\ \bibinfo {pages}
  {903} (\bibinfo {year} {2009})}\BibitemShut {NoStop}%
\bibitem [{\citenamefont {Bluhm}\ \emph
  {et~al.}(2010{\natexlab{b}})\citenamefont {Bluhm}, \citenamefont {Foletti},
  \citenamefont {Mahalu}, \citenamefont {Umansky},\ and\ \citenamefont
  {Yacoby}}]{Bluhm.10c}%
  \BibitemOpen
  \bibfield  {author} {\bibinfo {author} {\bibfnamefont {H.}~\bibnamefont
  {Bluhm}}, \bibinfo {author} {\bibfnamefont {S.}~\bibnamefont {Foletti}},
  \bibinfo {author} {\bibfnamefont {D.}~\bibnamefont {Mahalu}}, \bibinfo
  {author} {\bibfnamefont {V.}~\bibnamefont {Umansky}}, \ and\ \bibinfo
  {author} {\bibfnamefont {A.}~\bibnamefont {Yacoby}},\ }\href
  {https://link.aps.org/doi/10.1103/PhysRevLett.105.216803} {\bibfield
  {journal} {\bibinfo  {journal} {Phys. Rev. Lett.}\ }\textbf {\bibinfo
  {volume} {105}},\ \bibinfo {pages} {216803} (\bibinfo {year}
  {2010}{\natexlab{b}})}\BibitemShut {NoStop}%
\bibitem [{\citenamefont {van Weperen}\ \emph {et~al.}(2011)\citenamefont {van
  Weperen}, \citenamefont {Armstrong}, \citenamefont {Laird}, \citenamefont
  {Medford}, \citenamefont {Marcus}, \citenamefont {Hanson},\ and\
  \citenamefont {Gossard}}]{vanWeperen.11}%
  \BibitemOpen
  \bibfield  {author} {\bibinfo {author} {\bibfnamefont {I.}~\bibnamefont {van
  Weperen}}, \bibinfo {author} {\bibfnamefont {B.}~\bibnamefont {Armstrong}},
  \bibinfo {author} {\bibfnamefont {E.}~\bibnamefont {Laird}}, \bibinfo
  {author} {\bibfnamefont {J.}~\bibnamefont {Medford}}, \bibinfo {author}
  {\bibfnamefont {C.}~\bibnamefont {Marcus}}, \bibinfo {author} {\bibfnamefont
  {M.}~\bibnamefont {Hanson}}, \ and\ \bibinfo {author} {\bibfnamefont
  {A.}~\bibnamefont {Gossard}},\ }\href
  {https://link.aps.org/doi/10.1103/PhysRevLett.107.030506} {\bibfield
  {journal} {\bibinfo  {journal} {Phys. Rev. Lett.}\ }\textbf {\bibinfo
  {volume} {107}},\ \bibinfo {pages} {030506} (\bibinfo {year}
  {2011})}\BibitemShut {NoStop}%
\bibitem [{\citenamefont {Shulman}\ \emph {et~al.}(2012)\citenamefont
  {Shulman}, \citenamefont {Dial}, \citenamefont {Harvey}, \citenamefont
  {Bluhm}, \citenamefont {Umansky},\ and\ \citenamefont {Yacoby}}]{Shulman.12}%
  \BibitemOpen
  \bibfield  {author} {\bibinfo {author} {\bibfnamefont {M.~D.}\ \bibnamefont
  {Shulman}}, \bibinfo {author} {\bibfnamefont {O.~E.}\ \bibnamefont {Dial}},
  \bibinfo {author} {\bibfnamefont {S.~P.}\ \bibnamefont {Harvey}}, \bibinfo
  {author} {\bibfnamefont {H.}~\bibnamefont {Bluhm}}, \bibinfo {author}
  {\bibfnamefont {V.}~\bibnamefont {Umansky}}, \ and\ \bibinfo {author}
  {\bibfnamefont {A.}~\bibnamefont {Yacoby}},\ }\href
  {http://science.sciencemag.org/content/336/6078/202} {\bibfield  {journal}
  {\bibinfo  {journal} {Science}\ }\textbf {\bibinfo {volume} {336}},\ \bibinfo
  {pages} {202} (\bibinfo {year} {2012})}\BibitemShut {NoStop}%
\bibitem [{\citenamefont {Wu}\ \emph {et~al.}(2014)\citenamefont {Wu},
  \citenamefont {Ward}, \citenamefont {Prance}, \citenamefont {Kim},
  \citenamefont {Gamble}, \citenamefont {Mohr}, \citenamefont {Shi},
  \citenamefont {Savage}, \citenamefont {Lagally}, \citenamefont {Friesen},
  \citenamefont {Coppersmith},\ and\ \citenamefont {Eriksson}}]{Wu.14}%
  \BibitemOpen
  \bibfield  {author} {\bibinfo {author} {\bibfnamefont {X.}~\bibnamefont
  {Wu}}, \bibinfo {author} {\bibfnamefont {D.~R.}\ \bibnamefont {Ward}},
  \bibinfo {author} {\bibfnamefont {J.~R.}\ \bibnamefont {Prance}}, \bibinfo
  {author} {\bibfnamefont {D.}~\bibnamefont {Kim}}, \bibinfo {author}
  {\bibfnamefont {J.~K.}\ \bibnamefont {Gamble}}, \bibinfo {author}
  {\bibfnamefont {R.~T.}\ \bibnamefont {Mohr}}, \bibinfo {author}
  {\bibfnamefont {Z.}~\bibnamefont {Shi}}, \bibinfo {author} {\bibfnamefont
  {D.~E.}\ \bibnamefont {Savage}}, \bibinfo {author} {\bibfnamefont {M.~G.}\
  \bibnamefont {Lagally}}, \bibinfo {author} {\bibfnamefont {M.}~\bibnamefont
  {Friesen}}, \bibinfo {author} {\bibfnamefont {S.~N.}\ \bibnamefont
  {Coppersmith}}, \ and\ \bibinfo {author} {\bibfnamefont {M.~A.}\ \bibnamefont
  {Eriksson}},\ }\href {\doibase 10.1073/pnas.1412230111} {\bibfield  {journal}
  {\bibinfo  {journal} {Proc. Natl. Acad. Sci. U.S.A.}\ }\textbf {\bibinfo
  {volume} {111}},\ \bibinfo {pages} {11938} (\bibinfo {year}
  {2014})}\BibitemShut {NoStop}%
\bibitem [{\citenamefont {Nichol}\ \emph {et~al.}(2017)\citenamefont {Nichol},
  \citenamefont {Orona}, \citenamefont {Harvey}, \citenamefont {Fallahi},
  \citenamefont {Gardner}, \citenamefont {Manfra},\ and\ \citenamefont
  {Yacoby}}]{Nichol.16}%
  \BibitemOpen
  \bibfield  {author} {\bibinfo {author} {\bibfnamefont {J.~M.}\ \bibnamefont
  {Nichol}}, \bibinfo {author} {\bibfnamefont {L.~A.}\ \bibnamefont {Orona}},
  \bibinfo {author} {\bibfnamefont {S.~P.}\ \bibnamefont {Harvey}}, \bibinfo
  {author} {\bibfnamefont {S.}~\bibnamefont {Fallahi}}, \bibinfo {author}
  {\bibfnamefont {G.~C.}\ \bibnamefont {Gardner}}, \bibinfo {author}
  {\bibfnamefont {M.~J.}\ \bibnamefont {Manfra}}, \ and\ \bibinfo {author}
  {\bibfnamefont {A.}~\bibnamefont {Yacoby}},\ }\href
  {http://dx.doi.org/10.1038/s41534-016-0003-1} {\bibfield  {journal} {\bibinfo
   {journal} {npj Quantum Information}\ }\textbf {\bibinfo {volume} {3}},\
  \bibinfo {pages} {3} (\bibinfo {year} {2017})}\BibitemShut {NoStop}%
\bibitem [{\citenamefont {Reed}\ \emph {et~al.}(2016)\citenamefont {Reed},
  \citenamefont {Maune}, \citenamefont {Andrews}, \citenamefont {Borselli},
  \citenamefont {Eng}, \citenamefont {Jura}, \citenamefont {Kiselev},
  \citenamefont {Ladd}, \citenamefont {Merkel}, \citenamefont {Milosavljevic},
  \citenamefont {Pritchett}, \citenamefont {Rakher}, \citenamefont {Ross},
  \citenamefont {Schmitz}, \citenamefont {Smith}, \citenamefont {Wright},
  \citenamefont {Gyure},\ and\ \citenamefont {Hunter}}]{Reed.16}%
  \BibitemOpen
  \bibfield  {author} {\bibinfo {author} {\bibfnamefont {M.~D.}\ \bibnamefont
  {Reed}}, \bibinfo {author} {\bibfnamefont {B.~M.}\ \bibnamefont {Maune}},
  \bibinfo {author} {\bibfnamefont {R.~W.}\ \bibnamefont {Andrews}}, \bibinfo
  {author} {\bibfnamefont {M.~G.}\ \bibnamefont {Borselli}}, \bibinfo {author}
  {\bibfnamefont {K.}~\bibnamefont {Eng}}, \bibinfo {author} {\bibfnamefont
  {M.~P.}\ \bibnamefont {Jura}}, \bibinfo {author} {\bibfnamefont {A.~A.}\
  \bibnamefont {Kiselev}}, \bibinfo {author} {\bibfnamefont {T.~D.}\
  \bibnamefont {Ladd}}, \bibinfo {author} {\bibfnamefont {S.~T.}\ \bibnamefont
  {Merkel}}, \bibinfo {author} {\bibfnamefont {I.}~\bibnamefont
  {Milosavljevic}}, \bibinfo {author} {\bibfnamefont {E.~J.}\ \bibnamefont
  {Pritchett}}, \bibinfo {author} {\bibfnamefont {M.~T.}\ \bibnamefont
  {Rakher}}, \bibinfo {author} {\bibfnamefont {R.~S.}\ \bibnamefont {Ross}},
  \bibinfo {author} {\bibfnamefont {A.~E.}\ \bibnamefont {Schmitz}}, \bibinfo
  {author} {\bibfnamefont {A.}~\bibnamefont {Smith}}, \bibinfo {author}
  {\bibfnamefont {J.~A.}\ \bibnamefont {Wright}}, \bibinfo {author}
  {\bibfnamefont {M.~F.}\ \bibnamefont {Gyure}}, \ and\ \bibinfo {author}
  {\bibfnamefont {A.~T.}\ \bibnamefont {Hunter}},\ }\href
  {https://link.aps.org/doi/10.1103/PhysRevLett.116.110402} {\bibfield
  {journal} {\bibinfo  {journal} {Phys. Rev. Lett.}\ }\textbf {\bibinfo
  {volume} {116}},\ \bibinfo {pages} {110402} (\bibinfo {year}
  {2016})}\BibitemShut {NoStop}%
\bibitem [{\citenamefont {Martins}\ \emph {et~al.}(2016)\citenamefont
  {Martins}, \citenamefont {Malinowski}, \citenamefont {Nissen}, \citenamefont
  {Barnes}, \citenamefont {Fallahi}, \citenamefont {Gardner}, \citenamefont
  {Manfra}, \citenamefont {Marcus},\ and\ \citenamefont
  {Kuemmeth}}]{Martins.16}%
  \BibitemOpen
  \bibfield  {author} {\bibinfo {author} {\bibfnamefont {F.}~\bibnamefont
  {Martins}}, \bibinfo {author} {\bibfnamefont {F.~K.}\ \bibnamefont
  {Malinowski}}, \bibinfo {author} {\bibfnamefont {P.~D.}\ \bibnamefont
  {Nissen}}, \bibinfo {author} {\bibfnamefont {E.}~\bibnamefont {Barnes}},
  \bibinfo {author} {\bibfnamefont {S.}~\bibnamefont {Fallahi}}, \bibinfo
  {author} {\bibfnamefont {G.~C.}\ \bibnamefont {Gardner}}, \bibinfo {author}
  {\bibfnamefont {M.~J.}\ \bibnamefont {Manfra}}, \bibinfo {author}
  {\bibfnamefont {C.~M.}\ \bibnamefont {Marcus}}, \ and\ \bibinfo {author}
  {\bibfnamefont {F.}~\bibnamefont {Kuemmeth}},\ }\href
  {http://link.aps.org/doi/10.1103/PhysRevLett.116.116801} {\bibfield
  {journal} {\bibinfo  {journal} {Phys. Rev. Lett.}\ }\textbf {\bibinfo
  {volume} {116}},\ \bibinfo {pages} {116801} (\bibinfo {year}
  {2016})}\BibitemShut {NoStop}%
\bibitem [{\citenamefont {Hu}\ and\ \citenamefont {Das~Sarma}(2006)}]{Hu.06}%
  \BibitemOpen
  \bibfield  {author} {\bibinfo {author} {\bibfnamefont {X.}~\bibnamefont
  {Hu}}\ and\ \bibinfo {author} {\bibfnamefont {S.}~\bibnamefont {Das~Sarma}},\
  }\href {\doibase 10.1103/PhysRevLett.96.100501} {\bibfield  {journal}
  {\bibinfo  {journal} {Phys. Rev. Lett.}\ }\textbf {\bibinfo {volume} {96}},\
  \bibinfo {pages} {100501} (\bibinfo {year} {2006})}\BibitemShut {NoStop}%
\bibitem [{\citenamefont {Wardrop}\ and\ \citenamefont
  {Doherty}(2014)}]{Wardrop.14}%
  \BibitemOpen
  \bibfield  {author} {\bibinfo {author} {\bibfnamefont {M.~P.}\ \bibnamefont
  {Wardrop}}\ and\ \bibinfo {author} {\bibfnamefont {A.~C.}\ \bibnamefont
  {Doherty}},\ }\href {\doibase 10.1103/PhysRevB.90.045418} {\bibfield
  {journal} {\bibinfo  {journal} {Phys. Rev. B}\ }\textbf {\bibinfo {volume}
  {90}},\ \bibinfo {pages} {045418} (\bibinfo {year} {2014})}\BibitemShut
  {NoStop}%
\bibitem [{\citenamefont {{Thorgrimsson}}\ \emph {et~al.}(2017)\citenamefont
  {{Thorgrimsson}}, \citenamefont {{Kim}}, \citenamefont {{Yang}},
  \citenamefont {Smith}, \citenamefont {{Simmons}}, \citenamefont {{Ward}},
  \citenamefont {{Foote}}, \citenamefont {Corrigan}, \citenamefont {{Savage}},
  \citenamefont {{Lagally}}, \citenamefont {{Friesen}}, \citenamefont
  {{Coppersmith}},\ and\ \citenamefont {{Eriksson}}}]{Thorgrimsson.16}%
  \BibitemOpen
  \bibfield  {author} {\bibinfo {author} {\bibfnamefont {B.}~\bibnamefont
  {{Thorgrimsson}}}, \bibinfo {author} {\bibfnamefont {D.}~\bibnamefont
  {{Kim}}}, \bibinfo {author} {\bibfnamefont {Y.-C.}\ \bibnamefont {{Yang}}},
  \bibinfo {author} {\bibfnamefont {L.~W.}\ \bibnamefont {Smith}}, \bibinfo
  {author} {\bibfnamefont {C.~B.}\ \bibnamefont {{Simmons}}}, \bibinfo {author}
  {\bibfnamefont {D.~R.}\ \bibnamefont {{Ward}}}, \bibinfo {author}
  {\bibfnamefont {R.~H.}\ \bibnamefont {{Foote}}}, \bibinfo {author}
  {\bibfnamefont {J.}~\bibnamefont {Corrigan}}, \bibinfo {author}
  {\bibfnamefont {D.~E.}\ \bibnamefont {{Savage}}}, \bibinfo {author}
  {\bibfnamefont {M.~G.}\ \bibnamefont {{Lagally}}}, \bibinfo {author}
  {\bibfnamefont {M.}~\bibnamefont {{Friesen}}}, \bibinfo {author}
  {\bibfnamefont {S.~N.}\ \bibnamefont {{Coppersmith}}}, \ and\ \bibinfo
  {author} {\bibfnamefont {M.~A.}\ \bibnamefont {{Eriksson}}},\ }\href
  {https://www.nature.com/articles/s41534-017-0034-2} {\bibfield  {journal}
  {\bibinfo  {journal} {npj Quantum Information}\ }\textbf {\bibinfo {volume}
  {3}},\ \bibinfo {pages} {32} (\bibinfo {year} {2017})}\BibitemShut {NoStop}%
\bibitem [{\citenamefont {Yang}\ and\ \citenamefont
  {Wang}(2017{\natexlab{a}})}]{Yang.17b}%
  \BibitemOpen
  \bibfield  {author} {\bibinfo {author} {\bibfnamefont {X.-C.}\ \bibnamefont
  {Yang}}\ and\ \bibinfo {author} {\bibfnamefont {X.}~\bibnamefont {Wang}},\
  }\href {\doibase 10.1103/PhysRevA.96.012318} {\bibfield  {journal} {\bibinfo
  {journal} {Phys. Rev. A}\ }\textbf {\bibinfo {volume} {96}},\ \bibinfo
  {pages} {012318} (\bibinfo {year} {2017}{\natexlab{a}})}\BibitemShut
  {NoStop}%
\bibitem [{\citenamefont {Zhang}\ \emph {et~al.}(2017)\citenamefont {Zhang},
  \citenamefont {Throckmorton}, \citenamefont {Yang}, \citenamefont {Wang},
  \citenamefont {Barnes},\ and\ \citenamefont {Das~Sarma}}]{Zhang.17}%
  \BibitemOpen
  \bibfield  {author} {\bibinfo {author} {\bibfnamefont {C.}~\bibnamefont
  {Zhang}}, \bibinfo {author} {\bibfnamefont {R.~E.}\ \bibnamefont
  {Throckmorton}}, \bibinfo {author} {\bibfnamefont {X.-C.}\ \bibnamefont
  {Yang}}, \bibinfo {author} {\bibfnamefont {X.}~\bibnamefont {Wang}}, \bibinfo
  {author} {\bibfnamefont {E.}~\bibnamefont {Barnes}}, \ and\ \bibinfo {author}
  {\bibfnamefont {S.}~\bibnamefont {Das~Sarma}},\ }\href
  {https://link.aps.org/doi/10.1103/PhysRevLett.118.216802} {\bibfield
  {journal} {\bibinfo  {journal} {Phys. Rev. Lett.}\ }\textbf {\bibinfo
  {volume} {118}},\ \bibinfo {pages} {216802} (\bibinfo {year}
  {2017})}\BibitemShut {NoStop}%
\bibitem [{\citenamefont {Wood}\ and\ \citenamefont {Ellett}(1923)}]{Wood.23}%
  \BibitemOpen
  \bibfield  {author} {\bibinfo {author} {\bibfnamefont {R.}~\bibnamefont
  {Wood}}\ and\ \bibinfo {author} {\bibfnamefont {A.}~\bibnamefont {Ellett}},\
  }\href {http://dx.doi.org/10.1098/rspa.1923.0065} {\bibfield  {journal}
  {\bibinfo  {journal} {Proc. R. Soc. A}\ }\textbf {\bibinfo {volume} {103}},\
  \bibinfo {pages} {396} (\bibinfo {year} {1923})}\BibitemShut {NoStop}%
\bibitem [{\citenamefont {Eldridge}(1924)}]{Eldridge.24}%
  \BibitemOpen
  \bibfield  {author} {\bibinfo {author} {\bibfnamefont {J.~A.}\ \bibnamefont
  {Eldridge}},\ }\href
  {https://journals.aps.org/pr/abstract/10.1103/PhysRev.24.234} {\bibfield
  {journal} {\bibinfo  {journal} {Phys. Rev.}\ }\textbf {\bibinfo {volume}
  {24}},\ \bibinfo {pages} {234} (\bibinfo {year} {1924})}\BibitemShut
  {NoStop}%
\bibitem [{\citenamefont {Van~Vleck}(1925)}]{Van.25}%
  \BibitemOpen
  \bibfield  {author} {\bibinfo {author} {\bibfnamefont {J.~H.}\ \bibnamefont
  {Van~Vleck}},\ }\href {http://www.pnas.org/content/11/10/612.short}
  {\bibfield  {journal} {\bibinfo  {journal} {Proc. Natl. Acad. Sci. U.S.A.}\
  }\textbf {\bibinfo {volume} {11}},\ \bibinfo {pages} {612} (\bibinfo {year}
  {1925})}\BibitemShut {NoStop}%
\bibitem [{\citenamefont {Gunn}\ and\ \citenamefont {Sandle}(1971)}]{Gunn.71}%
  \BibitemOpen
  \bibfield  {author} {\bibinfo {author} {\bibfnamefont {H.~I.}\ \bibnamefont
  {Gunn}}\ and\ \bibinfo {author} {\bibfnamefont {W.~J.}\ \bibnamefont
  {Sandle}},\ }\href
  {http://iopscience.iop.org/article/10.1088/0022-3700/4/1/001/meta} {\bibfield
   {journal} {\bibinfo  {journal} {J. Phys. B}\ }\textbf {\bibinfo {volume}
  {4}},\ \bibinfo {pages} {L1} (\bibinfo {year} {1971})}\BibitemShut {NoStop}%
\bibitem [{\citenamefont {Deech}\ \emph {et~al.}(1975)\citenamefont {Deech},
  \citenamefont {Luypaert},\ and\ \citenamefont {Series}}]{Deech.75}%
  \BibitemOpen
  \bibfield  {author} {\bibinfo {author} {\bibfnamefont {J.~S.}\ \bibnamefont
  {Deech}}, \bibinfo {author} {\bibfnamefont {R.}~\bibnamefont {Luypaert}}, \
  and\ \bibinfo {author} {\bibfnamefont {G.~W.}\ \bibnamefont {Series}},\
  }\href {http://iopscience.iop.org/article/10.1088/0022-3700/8/9/007/meta}
  {\bibfield  {journal} {\bibinfo  {journal} {J. Phys. B}\ }\textbf {\bibinfo
  {volume} {8}},\ \bibinfo {pages} {1406} (\bibinfo {year} {1975})}\BibitemShut
  {NoStop}%
\bibitem [{\citenamefont {Sansonetti}\ \emph {et~al.}(2011)\citenamefont
  {Sansonetti}, \citenamefont {Simien}, \citenamefont {Gillaspy}, \citenamefont
  {Tan}, \citenamefont {Brewer}, \citenamefont {Brown}, \citenamefont {Wu},\
  and\ \citenamefont {Porto}}]{Sansonetti.11}%
  \BibitemOpen
  \bibfield  {author} {\bibinfo {author} {\bibfnamefont {C.~J.}\ \bibnamefont
  {Sansonetti}}, \bibinfo {author} {\bibfnamefont {C.~E.}\ \bibnamefont
  {Simien}}, \bibinfo {author} {\bibfnamefont {J.~D.}\ \bibnamefont
  {Gillaspy}}, \bibinfo {author} {\bibfnamefont {J.~N.}\ \bibnamefont {Tan}},
  \bibinfo {author} {\bibfnamefont {S.~M.}\ \bibnamefont {Brewer}}, \bibinfo
  {author} {\bibfnamefont {R.~C.}\ \bibnamefont {Brown}}, \bibinfo {author}
  {\bibfnamefont {S.}~\bibnamefont {Wu}}, \ and\ \bibinfo {author}
  {\bibfnamefont {J.~V.}\ \bibnamefont {Porto}},\ }\href
  {http://dx.doi.org/10.1103/PhysRevLett.107.023001} {\bibfield  {journal}
  {\bibinfo  {journal} {Phys. Rev. Lett.}\ }\textbf {\bibinfo {volume} {107}},\
  \bibinfo {pages} {023001} (\bibinfo {year} {2011})}\BibitemShut {NoStop}%
\bibitem [{\citenamefont {Lowe}(1959)}]{Lowe.59}%
  \BibitemOpen
  \bibfield  {author} {\bibinfo {author} {\bibfnamefont {I.~J.}\ \bibnamefont
  {Lowe}},\ }\href {\doibase 10.1103/PhysRevLett.2.285} {\bibfield  {journal}
  {\bibinfo  {journal} {Phys. Rev. Lett.}\ }\textbf {\bibinfo {volume} {2}},\
  \bibinfo {pages} {285} (\bibinfo {year} {1959})}\BibitemShut {NoStop}%
\bibitem [{\citenamefont {Andrew}\ \emph {et~al.}(1959)\citenamefont {Andrew},
  \citenamefont {Bradbury},\ and\ \citenamefont {Eades}}]{Andrew.59}%
  \BibitemOpen
  \bibfield  {author} {\bibinfo {author} {\bibfnamefont {E.~R.}\ \bibnamefont
  {Andrew}}, \bibinfo {author} {\bibfnamefont {A.}~\bibnamefont {Bradbury}}, \
  and\ \bibinfo {author} {\bibfnamefont {R.~G.}\ \bibnamefont {Eades}},\ }\href
  {http://dx.doi.org/10.1038/1831802a0} {\bibfield  {journal} {\bibinfo
  {journal} {Nature}\ }\textbf {\bibinfo {volume} {183}},\ \bibinfo {pages}
  {1802} (\bibinfo {year} {1959})}\BibitemShut {NoStop}%
\bibitem [{\citenamefont {Dreitlein}\ and\ \citenamefont
  {Kessemeier}(1961)}]{Dreitlein.61}%
  \BibitemOpen
  \bibfield  {author} {\bibinfo {author} {\bibfnamefont {J.}~\bibnamefont
  {Dreitlein}}\ and\ \bibinfo {author} {\bibfnamefont {H.}~\bibnamefont
  {Kessemeier}},\ }\href {\doibase 10.1103/PhysRev.123.835} {\bibfield
  {journal} {\bibinfo  {journal} {Phys. Rev.}\ }\textbf {\bibinfo {volume}
  {123}},\ \bibinfo {pages} {835} (\bibinfo {year} {1961})}\BibitemShut
  {NoStop}%
\bibitem [{\citenamefont {Kessemeier}\ and\ \citenamefont
  {Norberg}(1967)}]{Kessemeier.67}%
  \BibitemOpen
  \bibfield  {author} {\bibinfo {author} {\bibfnamefont {H.}~\bibnamefont
  {Kessemeier}}\ and\ \bibinfo {author} {\bibfnamefont {R.~E.}\ \bibnamefont
  {Norberg}},\ }\href {\doibase 10.1103/PhysRev.155.321} {\bibfield  {journal}
  {\bibinfo  {journal} {Phys. Rev.}\ }\textbf {\bibinfo {volume} {155}},\
  \bibinfo {pages} {321} (\bibinfo {year} {1967})}\BibitemShut {NoStop}%
\bibitem [{\citenamefont {Hennel}\ and\ \citenamefont
  {Klinowski}(2005)}]{Hennel.05}%
  \BibitemOpen
  \bibfield  {author} {\bibinfo {author} {\bibfnamefont {J.~W.}\ \bibnamefont
  {Hennel}}\ and\ \bibinfo {author} {\bibfnamefont {J.}~\bibnamefont
  {Klinowski}},\ }in\ \href {http://dx.doi.org/10.1007/b98646} {\emph {\bibinfo
  {booktitle} {New techniques in solid-state {NMR}}}}\ (\bibinfo  {publisher}
  {Springer},\ \bibinfo {year} {2005})\ pp.\ \bibinfo {pages}
  {1--14}\BibitemShut {NoStop}%
\bibitem [{\citenamefont {H{\"a}ndel}\ \emph {et~al.}(2003)\citenamefont
  {H{\"a}ndel}, \citenamefont {Gesele}, \citenamefont {Gottschall},\ and\
  \citenamefont {Albert}}]{Handel.03}%
  \BibitemOpen
  \bibfield  {author} {\bibinfo {author} {\bibfnamefont {H.}~\bibnamefont
  {H{\"a}ndel}}, \bibinfo {author} {\bibfnamefont {E.}~\bibnamefont {Gesele}},
  \bibinfo {author} {\bibfnamefont {K.}~\bibnamefont {Gottschall}}, \ and\
  \bibinfo {author} {\bibfnamefont {K.}~\bibnamefont {Albert}},\ }\href
  {\doibase 10.1002/anie.200390133} {\bibfield  {journal} {\bibinfo  {journal}
  {Angew. Chem. Int. Ed.}\ }\textbf {\bibinfo {volume} {42}},\ \bibinfo {pages}
  {438} (\bibinfo {year} {2003})}\BibitemShut {NoStop}%
\bibitem [{\citenamefont {Bydder}\ \emph {et~al.}(2007)\citenamefont {Bydder},
  \citenamefont {Rahal}, \citenamefont {Fullerton},\ and\ \citenamefont
  {Bydder}}]{Bydder.07}%
  \BibitemOpen
  \bibfield  {author} {\bibinfo {author} {\bibfnamefont {M.}~\bibnamefont
  {Bydder}}, \bibinfo {author} {\bibfnamefont {A.}~\bibnamefont {Rahal}},
  \bibinfo {author} {\bibfnamefont {G.~D.}\ \bibnamefont {Fullerton}}, \ and\
  \bibinfo {author} {\bibfnamefont {G.~M.}\ \bibnamefont {Bydder}},\ }\href
  {http://dx.doi.org/10.1002/jmri.20850} {\bibfield  {journal} {\bibinfo
  {journal} {J. Magn. Reson. Imaging}\ }\textbf {\bibinfo {volume} {25}},\
  \bibinfo {pages} {290} (\bibinfo {year} {2007})}\BibitemShut {NoStop}%
\bibitem [{\citenamefont {Ding}\ \emph {et~al.}(2001)\citenamefont {Ding},
  \citenamefont {McDowell}, \citenamefont {Ye}, \citenamefont {Zhan},
  \citenamefont {Zhu}, \citenamefont {Gao}, \citenamefont {Sun}, \citenamefont
  {Mao},\ and\ \citenamefont {Liu}}]{Ding.01}%
  \BibitemOpen
  \bibfield  {author} {\bibinfo {author} {\bibfnamefont {S.}~\bibnamefont
  {Ding}}, \bibinfo {author} {\bibfnamefont {C.~A.}\ \bibnamefont {McDowell}},
  \bibinfo {author} {\bibfnamefont {C.}~\bibnamefont {Ye}}, \bibinfo {author}
  {\bibfnamefont {M.}~\bibnamefont {Zhan}}, \bibinfo {author} {\bibfnamefont
  {X.}~\bibnamefont {Zhu}}, \bibinfo {author} {\bibfnamefont {K.}~\bibnamefont
  {Gao}}, \bibinfo {author} {\bibfnamefont {X.}~\bibnamefont {Sun}}, \bibinfo
  {author} {\bibfnamefont {X.-A.}\ \bibnamefont {Mao}}, \ and\ \bibinfo
  {author} {\bibfnamefont {M.}~\bibnamefont {Liu}},\ }\href
  {http://link.springer.com/article/10.1007/s100510170018} {\bibfield
  {journal} {\bibinfo  {journal} {Eur. Phys. J. B}\ }\textbf {\bibinfo {volume}
  {24}},\ \bibinfo {pages} {23} (\bibinfo {year} {2001})}\BibitemShut {NoStop}%
\bibitem [{\citenamefont {Lu}\ \emph {et~al.}(2009)\citenamefont {Lu},
  \citenamefont {Zhou}, \citenamefont {Kuang},\ and\ \citenamefont
  {Sun}}]{Lu.09}%
  \BibitemOpen
  \bibfield  {author} {\bibinfo {author} {\bibfnamefont {J.}~\bibnamefont
  {Lu}}, \bibinfo {author} {\bibfnamefont {L.}~\bibnamefont {Zhou}}, \bibinfo
  {author} {\bibfnamefont {L.-M.}\ \bibnamefont {Kuang}}, \ and\ \bibinfo
  {author} {\bibfnamefont {C.~P.}\ \bibnamefont {Sun}},\ }\href
  {http://dx.doi.org/10.1103/PhysRevE.79.016606} {\bibfield  {journal}
  {\bibinfo  {journal} {Phys. Rev. E}\ }\textbf {\bibinfo {volume} {79}},\
  \bibinfo {pages} {016606} (\bibinfo {year} {2009})}\BibitemShut {NoStop}%
\bibitem [{pol()}]{polar}%
  \BibitemOpen
  \href@noop {} {}\bibinfo {note} {The polar axis is defined as the one
  penetrating the center of both dots.}\BibitemShut {Stop}%
\bibitem [{\citenamefont {Burkard}\ \emph {et~al.}(1999)\citenamefont
  {Burkard}, \citenamefont {Loss},\ and\ \citenamefont
  {DiVincenzo}}]{Burkard.99}%
  \BibitemOpen
  \bibfield  {author} {\bibinfo {author} {\bibfnamefont {G.}~\bibnamefont
  {Burkard}}, \bibinfo {author} {\bibfnamefont {D.}~\bibnamefont {Loss}}, \
  and\ \bibinfo {author} {\bibfnamefont {D.~P.}\ \bibnamefont {DiVincenzo}},\
  }\href {\doibase 10.1103/PhysRevB.59.2070} {\bibfield  {journal} {\bibinfo
  {journal} {Phys. Rev. B}\ }\textbf {\bibinfo {volume} {59}},\ \bibinfo
  {pages} {2070} (\bibinfo {year} {1999})}\BibitemShut {NoStop}%
\bibitem [{\citenamefont {Yang}\ \emph {et~al.}(2011)\citenamefont {Yang},
  \citenamefont {Wang},\ and\ \citenamefont {Das~Sarma}}]{Yang.11}%
  \BibitemOpen
  \bibfield  {author} {\bibinfo {author} {\bibfnamefont {S.}~\bibnamefont
  {Yang}}, \bibinfo {author} {\bibfnamefont {X.}~\bibnamefont {Wang}}, \ and\
  \bibinfo {author} {\bibfnamefont {S.}~\bibnamefont {Das~Sarma}},\ }\href
  {\doibase 10.1103/PhysRevB.83.161301} {\bibfield  {journal} {\bibinfo
  {journal} {Phys. Rev. B}\ }\textbf {\bibinfo {volume} {83}},\ \bibinfo
  {pages} {161301} (\bibinfo {year} {2011})}\BibitemShut {NoStop}%
\bibitem [{\citenamefont {Wang}\ \emph {et~al.}(2011)\citenamefont {Wang},
  \citenamefont {Yang},\ and\ \citenamefont {Das~Sarma}}]{Wang.11}%
  \BibitemOpen
  \bibfield  {author} {\bibinfo {author} {\bibfnamefont {X.}~\bibnamefont
  {Wang}}, \bibinfo {author} {\bibfnamefont {S.}~\bibnamefont {Yang}}, \ and\
  \bibinfo {author} {\bibfnamefont {S.}~\bibnamefont {Das~Sarma}},\ }\href
  {\doibase 10.1103/PhysRevB.84.115301} {\bibfield  {journal} {\bibinfo
  {journal} {Phys. Rev. B}\ }\textbf {\bibinfo {volume} {84}},\ \bibinfo
  {pages} {115301} (\bibinfo {year} {2011})}\BibitemShut {NoStop}%
\bibitem [{\citenamefont {Li}\ \emph {et~al.}(2012)\citenamefont {Li},
  \citenamefont {Hu},\ and\ \citenamefont {You}}]{Li.12}%
  \BibitemOpen
  \bibfield  {author} {\bibinfo {author} {\bibfnamefont {R.}~\bibnamefont
  {Li}}, \bibinfo {author} {\bibfnamefont {X.}~\bibnamefont {Hu}}, \ and\
  \bibinfo {author} {\bibfnamefont {J.~Q.}\ \bibnamefont {You}},\ }\href
  {\doibase 10.1103/PhysRevB.86.205306} {\bibfield  {journal} {\bibinfo
  {journal} {Phys. Rev. B}\ }\textbf {\bibinfo {volume} {86}},\ \bibinfo
  {pages} {205306} (\bibinfo {year} {2012})}\BibitemShut {NoStop}%
\bibitem [{\citenamefont {Yang}\ and\ \citenamefont
  {Wang}(2017{\natexlab{b}})}]{Yang.17a}%
  \BibitemOpen
  \bibfield  {author} {\bibinfo {author} {\bibfnamefont {X.-C.}\ \bibnamefont
  {Yang}}\ and\ \bibinfo {author} {\bibfnamefont {X.}~\bibnamefont {Wang}},\
  }\href {\doibase 10.1103/PhysRevA.95.052325} {\bibfield  {journal} {\bibinfo
  {journal} {Phys. Rev. A}\ }\textbf {\bibinfo {volume} {95}},\ \bibinfo
  {pages} {052325} (\bibinfo {year} {2017}{\natexlab{b}})}\BibitemShut
  {NoStop}%
\bibitem [{\citenamefont {Simmons}\ \emph {et~al.}(2009)\citenamefont
  {Simmons}, \citenamefont {Thalakulam}, \citenamefont {Rosemeyer},
  \citenamefont {Van~Bael}, \citenamefont {Sackmann}, \citenamefont {Savage},
  \citenamefont {Lagally}, \citenamefont {Joynt}, \citenamefont {Friesen},
  \citenamefont {Coppersmith},\ and\ \citenamefont {Eriksson}}]{Simmons.09}%
  \BibitemOpen
  \bibfield  {author} {\bibinfo {author} {\bibfnamefont {C.~B.}\ \bibnamefont
  {Simmons}}, \bibinfo {author} {\bibfnamefont {M.}~\bibnamefont {Thalakulam}},
  \bibinfo {author} {\bibfnamefont {B.~M.}\ \bibnamefont {Rosemeyer}}, \bibinfo
  {author} {\bibfnamefont {B.~J.}\ \bibnamefont {Van~Bael}}, \bibinfo {author}
  {\bibfnamefont {E.~K.}\ \bibnamefont {Sackmann}}, \bibinfo {author}
  {\bibfnamefont {D.~E.}\ \bibnamefont {Savage}}, \bibinfo {author}
  {\bibfnamefont {M.~G.}\ \bibnamefont {Lagally}}, \bibinfo {author}
  {\bibfnamefont {R.}~\bibnamefont {Joynt}}, \bibinfo {author} {\bibfnamefont
  {M.}~\bibnamefont {Friesen}}, \bibinfo {author} {\bibfnamefont {S.~N.}\
  \bibnamefont {Coppersmith}}, \ and\ \bibinfo {author} {\bibfnamefont {M.~A.}\
  \bibnamefont {Eriksson}},\ }\href {\doibase 10.1021/nl9014974} {\bibfield
  {journal} {\bibinfo  {journal} {Nano Lett.}\ }\textbf {\bibinfo {volume}
  {9}},\ \bibinfo {pages} {3234} (\bibinfo {year} {2009})}\BibitemShut
  {NoStop}%
\bibitem [{\citenamefont {Zajac}\ \emph {et~al.}(2016)\citenamefont {Zajac},
  \citenamefont {Hazard}, \citenamefont {Mi}, \citenamefont {Nielsen},\ and\
  \citenamefont {Petta}}]{Zajac.16}%
  \BibitemOpen
  \bibfield  {author} {\bibinfo {author} {\bibfnamefont {D.~M.}\ \bibnamefont
  {Zajac}}, \bibinfo {author} {\bibfnamefont {T.~M.}\ \bibnamefont {Hazard}},
  \bibinfo {author} {\bibfnamefont {X.}~\bibnamefont {Mi}}, \bibinfo {author}
  {\bibfnamefont {E.}~\bibnamefont {Nielsen}}, \ and\ \bibinfo {author}
  {\bibfnamefont {J.~R.}\ \bibnamefont {Petta}},\ }\href
  {http://link.aps.org/doi/10.1103/PhysRevApplied.6.054013} {\bibfield
  {journal} {\bibinfo  {journal} {Phys. Rev. Applied}\ }\textbf {\bibinfo
  {volume} {6}},\ \bibinfo {pages} {054013} (\bibinfo {year}
  {2016})}\BibitemShut {NoStop}%
\bibitem [{\citenamefont {Jones}\ \emph {et~al.}(2016)\citenamefont {Jones},
  \citenamefont {Gyure}, \citenamefont {Ladd}, \citenamefont {Fogarty},
  \citenamefont {Morello},\ and\ \citenamefont {Dzurak}}]{Jones.16}%
  \BibitemOpen
  \bibfield  {author} {\bibinfo {author} {\bibfnamefont {C.}~\bibnamefont
  {Jones}}, \bibinfo {author} {\bibfnamefont {M.~F.}\ \bibnamefont {Gyure}},
  \bibinfo {author} {\bibfnamefont {T.~D.}\ \bibnamefont {Ladd}}, \bibinfo
  {author} {\bibfnamefont {M.~A.}\ \bibnamefont {Fogarty}}, \bibinfo {author}
  {\bibfnamefont {A.}~\bibnamefont {Morello}}, \ and\ \bibinfo {author}
  {\bibfnamefont {A.~S.}\ \bibnamefont {Dzurak}},\ }\href
  {https://arxiv.org/abs/1608.06335} {\bibfield  {journal} {\bibinfo  {journal}
  {arXiv preprint arXiv:1608.06335}\ } (\bibinfo {year} {2016})}\BibitemShut
  {NoStop}%
\bibitem [{\citenamefont {Trifunovic}\ \emph {et~al.}(2012)\citenamefont
  {Trifunovic}, \citenamefont {Dial}, \citenamefont {Trif}, \citenamefont
  {Wootton}, \citenamefont {Abebe}, \citenamefont {Yacoby},\ and\ \citenamefont
  {Loss}}]{Trifunovic.12}%
  \BibitemOpen
  \bibfield  {author} {\bibinfo {author} {\bibfnamefont {L.}~\bibnamefont
  {Trifunovic}}, \bibinfo {author} {\bibfnamefont {O.}~\bibnamefont {Dial}},
  \bibinfo {author} {\bibfnamefont {M.}~\bibnamefont {Trif}}, \bibinfo {author}
  {\bibfnamefont {J.~R.}\ \bibnamefont {Wootton}}, \bibinfo {author}
  {\bibfnamefont {R.}~\bibnamefont {Abebe}}, \bibinfo {author} {\bibfnamefont
  {A.}~\bibnamefont {Yacoby}}, \ and\ \bibinfo {author} {\bibfnamefont
  {D.}~\bibnamefont {Loss}},\ }\href {\doibase 10.1103/PhysRevX.2.011006}
  {\bibfield  {journal} {\bibinfo  {journal} {Phys. Rev. X}\ }\textbf {\bibinfo
  {volume} {2}},\ \bibinfo {pages} {011006} (\bibinfo {year}
  {2012})}\BibitemShut {NoStop}%
\bibitem [{\citenamefont {Oosterkamp}\ \emph {et~al.}(1998)\citenamefont
  {Oosterkamp}, \citenamefont {Fujisawa}, \citenamefont {Van Der~Wiel},
  \citenamefont {Ishibashi}, \citenamefont {Hijman}, \citenamefont {Tarucha},\
  and\ \citenamefont {Kouwenhoven}}]{Oosterkamp.98}%
  \BibitemOpen
  \bibfield  {author} {\bibinfo {author} {\bibfnamefont {T.~H.}\ \bibnamefont
  {Oosterkamp}}, \bibinfo {author} {\bibfnamefont {T.}~\bibnamefont
  {Fujisawa}}, \bibinfo {author} {\bibfnamefont {W.~G.}\ \bibnamefont {Van
  Der~Wiel}}, \bibinfo {author} {\bibfnamefont {K.}~\bibnamefont {Ishibashi}},
  \bibinfo {author} {\bibfnamefont {R.~V.}\ \bibnamefont {Hijman}}, \bibinfo
  {author} {\bibfnamefont {S.}~\bibnamefont {Tarucha}}, \ and\ \bibinfo
  {author} {\bibfnamefont {L.~P.}\ \bibnamefont {Kouwenhoven}},\ }\href
  {http://www.nature.com/nature/journal/v395/n6705/abs/395873a0.html}
  {\bibfield  {journal} {\bibinfo  {journal} {Nature}\ }\textbf {\bibinfo
  {volume} {395}},\ \bibinfo {pages} {873} (\bibinfo {year}
  {1998})}\BibitemShut {NoStop}%
\bibitem [{\citenamefont {Cao}\ \emph {et~al.}(2016)\citenamefont {Cao},
  \citenamefont {Li}, \citenamefont {Yu}, \citenamefont {Wang}, \citenamefont
  {Chen}, \citenamefont {Song}, \citenamefont {Xiao}, \citenamefont {Guo},
  \citenamefont {Jiang}, \citenamefont {Hu},\ and\ \citenamefont
  {Guo}}]{Cao.16}%
  \BibitemOpen
  \bibfield  {author} {\bibinfo {author} {\bibfnamefont {G.}~\bibnamefont
  {Cao}}, \bibinfo {author} {\bibfnamefont {H.-O.}\ \bibnamefont {Li}},
  \bibinfo {author} {\bibfnamefont {G.-D.}\ \bibnamefont {Yu}}, \bibinfo
  {author} {\bibfnamefont {B.-C.}\ \bibnamefont {Wang}}, \bibinfo {author}
  {\bibfnamefont {B.-B.}\ \bibnamefont {Chen}}, \bibinfo {author}
  {\bibfnamefont {X.-X.}\ \bibnamefont {Song}}, \bibinfo {author}
  {\bibfnamefont {M.}~\bibnamefont {Xiao}}, \bibinfo {author} {\bibfnamefont
  {G.-C.}\ \bibnamefont {Guo}}, \bibinfo {author} {\bibfnamefont {H.-W.}\
  \bibnamefont {Jiang}}, \bibinfo {author} {\bibfnamefont {X.}~\bibnamefont
  {Hu}}, \ and\ \bibinfo {author} {\bibfnamefont {G.-P.}\ \bibnamefont {Guo}},\
  }\href {\doibase 10.1103/PhysRevLett.116.086801} {\bibfield  {journal}
  {\bibinfo  {journal} {Phys. Rev. Lett.}\ }\textbf {\bibinfo {volume} {116}},\
  \bibinfo {pages} {086801} (\bibinfo {year} {2016})}\BibitemShut {NoStop}%
\end{thebibliography}

%

\end{document}